\documentclass[aps,pre, superscriptaddress, reprint, amsmath, amssymb]{revtex4-2}
\usepackage{graphicx} 
\usepackage[dvipsnames]{xcolor}
\usepackage{amsmath,amssymb}
\usepackage{esint}
\usepackage{physics}
\usepackage{float}
\usepackage[colorlinks]{hyperref}
\hypersetup{
	linkcolor= NavyBlue,
	citecolor=	ForestGreen,
	linktocpage=true}
\graphicspath{ {./Figs/} }

\begin{document}
\title{Correlated decoherence and thermometry with mobile impurities in a 1D Fermi gas}

\author{Sindre Brattegard} 
\email{brattegs@tcd.ie}
\affiliation{School of Physics, Trinity College Dublin, College Green, Dublin 2, Ireland}
\author{Thom\'as Fogarty}
\affiliation{Quantum Systems Unit, OIST Graduate University, Onna, Okinawa 904-0495, Japan}
\author{Thomas Busch}
\affiliation{Quantum Systems Unit, OIST Graduate University, Onna, Okinawa 904-0495, Japan}
\author{Mark T. Mitchison}
\email{mark.mitchison@kcl.ac.uk}
\affiliation{School of Physics, Trinity College Dublin, College Green, Dublin 2, Ireland}
\affiliation{Department of Physics, King’s College London, Strand, London, WC2R 2LS, United Kingdom}

\begin{abstract}
We theoretically investigate the correlated decoherence dynamics of two mobile impurities trapped within a gas of ultracold fermionic atoms. We use a mean-field approximation to self-consistently describe the effect of impurity-gas collisions on impurity motion, while decoherence of the impurities' internal state is computed exactly within a functional determinant approach. At equilibrium, we find that the impurities undergo bath-induced localisation as the impurity-gas interaction strength is increased. We then study the non-equilibrium dynamics induced by a sudden change of the impurities' internal state, which can be experimentally probed by Ramsey interferometry. Our theoretical approach allows us to investigate the effect of impurity motion on decoherence dynamics, finding strong deviations from the universal behaviour associated with Anderson's orthogonality catastrophe when the mass imbalance between impurity and gas atoms is small. Finally, we show that mobile impurities can be used as thermometers of their environment and that bath-mediated correlations can be beneficial for thermometric performance at low temperatures, even in the presence of non-trivial impurity motion. Our results showcase the interesting open quantum dynamics of mobile impurities dephasing in a common environment, and could help provide more precise temperature estimates of ultracold fermionic mixtures.

\end{abstract}

\maketitle

\section{Introduction}
Modern experimental techniques for cold atomic gases allow for the precise control of confining potentials and interatomic interactions, making them an ideal testbed for many-body physics~\cite{BLOCH_2008}. In particular, two or more atomic species can be loaded into the same trap allowing to realize multicomponent systems, while species dependent traps can confine particles that are in different hyperfine levels into independent potentials with distinct geometries~\cite{sowinski_one-dimensional2019,recati_coherently_2022,  baroni_quantum_2024}. By making the density of one of the species much lower than the other they can be considered as impurities within a fluid of the majority atoms. Mobile impurities can get dressed by their environment, forming polaronic quasi-particles~\cite{massignan2025} in bosonic~\cite {yegovtsev_effective_2023, skou_non-equilibrium_2021,grusdt2024} and fermionic~\cite{Schirotzek2009,massignan_polarons_2014,Scazza2017, ness_observation_2020} environments. On the other hand, static impurities in a fermionic environment have been shown to exhibit universal behavior predicted by the Anderson orthogonality catastrophe (OC)~\cite{anderson_infrared_1967, goold_orthogonality_2011,knap_time-dependent_2012,cetina_decoherence_2015,cetina_ultrafast_2016}. Since typical ultracold atomic experiments operate with hundreds, or even thousands, of impurities, a substantial research effort has been put into understanding interactions between impurities mediated by the bath~\cite{Paredes2024,baroni_mediated_2024} and have even been observed experimentally~\cite{edri_observation_2020,baroni_mediated_2023_nat}. 

However, at ultralow temperatures, measuring the temperature of ultracold mixtures becomes difficult. Standard techniques based on time-of-flight completely destroy the system, lack spatial resolution, and may yield low precision at ultralow temperatures~\cite{onofrio_physics_2016}. Recently, there have been improvements in other techniques, such as thermometry based on density fluctuations, which have shown great potential~\cite{muller2010local,zhou2011universal,hartke2020doublon}. Despite such thermometry schemes being in principle non-destructive, and able to be extended to access the local temperature~\cite{dixmerias2025}, they may still suffer from low precision in the degenerate regime. Over the last decade, the study of the fundamental limits of temperature estimation in the quantum regime has gained prominence in the literature~\cite{mehboudi_thermometry_2019}. Early work in the field studied probes in equilibrium with a bath~\cite{correa_individual_2015}, but such setups always suffer from exponentially vanishing precision as temperature is lowered. Instead, several proposals have been put forward to use the non-equilibrium dynamics of a probe for thermometry, especially for ultracold gases~\cite{Bruderer_2006,sabin_impurities_2014, hangleiter_non-destructive_2015, Johnson2016, razavian_quantum_2019, mitchison_situ_2020,oghittu_quantum-limited_2022,mitchison_taking_2022,yuan_quantum_2023,ravell_rodriguez_strongly_2024}. Other important work has shown how thermometry precision is affected by strong system-bath correlations~\cite{correa_enhancement_2017,miller_energy-temperature_2018,mihailescu_thermometry_2023} and informational constraints, such as lack of knowledge of all relevant parameters~\cite{mihailescu2025metrologicalsymmetriessingularquantum}, limited measurement data~\cite{rubio_global_2021, mok_optimal_2021,boeyens_uninformed_2021, jorgensen_bayesian_2022, mehboudi_fundamental_2022,alves_bayesian_2022}, and coarse grained measurements~\cite{jorgensen_tight_2020,hovhannisyan_optimal_2021}. 

Another important physical effect that can be exploited in thermometry is the correlations of multiple probes due to their common thermal environment~\cite{planella_bath-induced_2022, brenes_multi-spin_2023}. In particular, it has been shown that, at low temperature and weak coupling, bath-induced correlations developing in time can be used as a resource to enhance thermometric precision~\cite{planella_bath-induced_2022, brenes_multi-spin_2023,gebbia_two-qubit_2020, Xu2025, Aiache2024, brattegard2024}. In these studies, the impurities are treated as static, distinguishable particles, and after an appropriate interaction time, the temperature is infered from a measurement on the internal state of the impurities. However, in experiments, some motion of the impurities is unavoidable. It is well-known that impurity motion affects the decoherence dynamics of the impurities, e.g. for impurities in a Fermi gas, the motion of impurities changes the universal OC dynamics. An understanding of how impurity motion affects decoherence dynamics, and how whether the collective thermometric precision enhancement prevails in the presence of impurity motion, is lacking in the literature.

In the following, we fill this gap by investigating mobile impurities embedded in a 1D homogeneous gas of fermions at finite temperature trapped in a box potential. Such box potentials can routinely be created in experimental labs~\cite{gaunt_bose-einstein_2013,schmidutz_quantum_2014, navon_critical_2015, mukherjee_homogeneous_2017}. In Sec.~\ref{description} we describe our setup of mobile impurities with two internal energy states and a spatial distribution. We describe this setup in detail in Sec.~\ref{description} and calculate the equilibrium state of the impurities and background gas based on a mean-field (MF) approach in Sec.~\ref{gs_calc}. We find that as the interaction between impurities and background gas increases, the impurities can start to localize while minimally perturbing the background gas. Importantly, as this happens, a gap opens up in the single-particle spectrum of the impurities which allows us to neglect the temperature of the impurities as long as the gap is much larger than the temperature of the background gas. This also allows us to neglect thermalization of the impurities with the environment and therefore only focus on how pure dephasing of the impurities is affected by temperature. Furthermore, we always take the impurities to be heavier than the majority gas atoms. 

Next, in Sec. \ref{quench_dyn}, we calculate the quench dynamics of the densities of the impurities and majority atoms if the internal state of one or more of the impurities is flipped. Since the internal states can couple differently to the background gas, this is effectively a quench in the interaction strengths of the impurities that have their internal state flipped. The interaction strength of each internal state can be tuned by the use of Feshbach resonances~\cite{chin_feshbach_2010}. We find that quenching the interaction strength between impurities and background gas pushes the majority atoms away from the impurity positions, leading to excitations travelling through the gas with the Fermi velocity. The density of the impurities, on the other hand, exhibits breathing modes~\cite{siegl_many-body_2018,huang_breathing_2019,Mistakidis2019,Mukherjee2020}. The excitations travelling through the gas give rise to bath-mediated interactions developing in real time. 

In Sec. \ref{decoherence_dyn}, we study the internal state of the impurities, which undergoes pure decoherence due to the absence of spin-changing collisions. Based on a functional determinant approach~(FDA)~\cite{abanin_fermi-edge_2005,dambrumenil_fermi_2005,schonhammer_full_2007,wang_exact_2022,wang_heavy_2022,wang_multidimensional_2023}, we calculate the decoherence dynamics of two impurities and show that for sufficient mass imbalance between the impurities and the host gas, the universal time-evolution of the decoherence functions previously studied for static impurities arises. However, our theoretical method is not limited to this regime and we also investigate the decoherence function for light impurities using a mean-field description of the impurity motion. This approach neglects quantum correlations between motional degrees of freedom and is thus limited to moderately weak impurity-gas coupling. However, the FDA provides a non-perturbative description of the spin- and motion-dependent decoherence functions, and the corresponding classical correlations between the spin, the impurity position, and the gas atoms.

Finally, in Sec. \ref{thermometry}, based on local quantum parameter estimation theory~\cite{paris_quantum_2009}, we show how the mobile impurities can be used as thermometers for the host gas. Mobile impurity thermometers have been suggested in other works, but then only by considering the equilibrium density of the impurities~\cite{wasak2025bosepolaronthermometertrapped,mehboudi_using_2019}. We find that for a single impurity, most of the temperature information is contained in the internal state of the impurity. We then consider a two-impurity thermometer, and find that at low temperature, and relatively small mass imbalance, there is a substantial enhancement of thermometric performance when comparing two correlated impurities to that of two independent impurities. Importantly, we show that this enhancement prevails even if we only have access to the internal state of the impurities, thus showing that the collective enhancement found for static impurities prevails even when the motion of impurities is taken into account. 

\section{Description of system}
\label{description}
We consider a system $S$ of $N_S$ fermionic impurity atoms (we will focus on $N_S=1,2$) interacting with an environment $E$ of $N_E \gg 1$ majority atoms. We take $E$ to be a gas of spin-polarized, non-interacting fermions confined to a 1D box potential of length $L$. This is a good description as long as the gas is dilute and the temperature $T$ is low enough for $s$-wave scattering to be dominant, because $s$-wave scattering between identical fermions is forbidden due to the anti-symmetry of fermionic wavefunctions.

The impurity atoms are modelled as two-level systems with energy eigenstates $|\sigma\rangle_n$ where $\sigma\in\{\downarrow,\uparrow\}$ and $n = 1,\ldots,N_S$ labels the different impurity atoms. The total Hamiltonian is given by
\begin{align}
\label{H_tot}
   & \hat H = \hat H_S + \hat H_E + \hat H_{I},\\
   \label{H_E}
   & \hat{H}_E  =  -\frac{\hbar^2}{2M_E}\int_{-L/2}^{L/2}\dd x \hat\Psi^\dagger(x)\partial_x^2\hat\Psi(x),
\end{align} 
where $M_E$ is the mass of the majority atoms and $\hat\Psi^\dagger(x)$ $(\hat\Psi(x))$ creates (destroys) a fermion at position $x$ and obeys the fermionic anti-commutation relation $\{\hat{\Psi}(x), \hat\Psi^\dagger(x^\prime)\} = \delta(x-x^\prime)$. The system Hamiltonian has a contribution from the internal qubit states as well as the kinetic and potential energy of the motion of the impurities in a trap:
\begin{align}
   & \hat H_{S} = \hat{H}_{\text{qubit}} + \hat{H}_{K} + \hat H_P ,\\ 
   & \hat{H}_{\text{qubit}} = \varepsilon\sum_{n=1}^{N_S}\ket{\uparrow}_n \!\bra{\uparrow},\\
   &\hat{H}_{K} =  -\frac{\hbar^2}{2M_S}\sum_{\sigma=\downarrow,\uparrow}\int_{-L/2}^{L/2}\dd x \hat\Phi_\sigma^\dagger(x)\partial_x^2\hat\Phi_\sigma(x),\\
   &\hat H_P = \sum_{\sigma=\downarrow,\uparrow}\int_{-L/2}^{L/2}\dd x  V(x) \hat\Phi_\sigma^\dagger(x)\hat\Phi_\sigma(x),
\end{align}
where $M_S$ is the impurity mass, $\varepsilon$ is the local energy splitting of the internal state of the impurities, and $\hat\Phi_\sigma^\dagger(x)$ $(\hat\Phi_\sigma)$ creates (destroys) an impurity fermion in internal state $\sigma$ at position $x$ and obeys the anti-commutation relation $\{\hat{\Phi}_\sigma(x), \hat\Phi^\dagger_{\sigma^\prime}(x^\prime)\} = \delta(x-x^\prime)\delta_{\sigma,\sigma^\prime}$. $V(x)$ is a species-selective confining potential that is only applied to the impurities. We take the potential to be a harmonic trap $V(x) = M_S\omega^2 x^2/2$, where $\omega$ is the frequency of the trap. The interaction between $S$ and $E$ is taken to be a contact interaction  
\begin{align}
\label{H_I}
    \hat{H}_{I} = \sum_{\sigma=\uparrow,\downarrow}\kappa_\sigma\int_{-L/2}^{L/2}\dd x\hat{\Psi}^\dagger(x)\hat{\Psi}(x)\hat{\Phi}_\sigma^\dagger(x)\hat{\Phi}_\sigma(x),
\end{align}
with a coupling strength $\kappa_\sigma$ that depends on the internal state of the impurity atoms.   

\subsection{Equilibrium state calculation}
\label{gs_calc}
\begin{figure}
    \centering
    \includegraphics[width=0.499\textwidth]{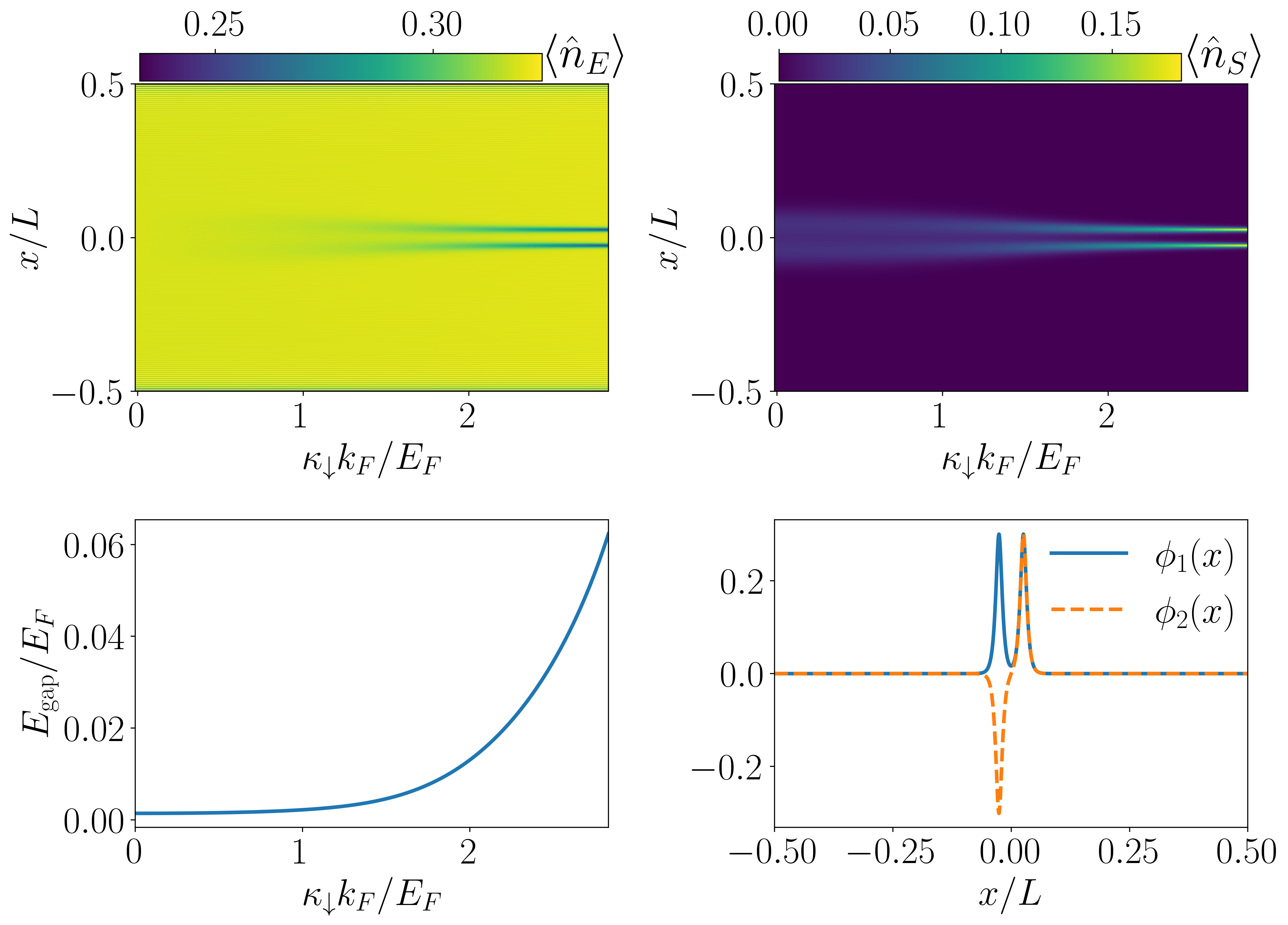}
    \caption{(a) and (b): The equilibrium density of the environment and impurities as the coupling strength is increased. (c) The energy gap from the second to the third energy eigenvalues of the Hamiltonian Eq.~(\ref{h_S}). (d) The single particle wavefunctions $\phi_1(x)$ and $\phi_2(x)$ at $\kappa_\downarrow k_F/E_F \approx 2.6$ In all cases we have $N_E = 150$, $T/T_F = 0.005$, $M_S = 2 M_E$, and $\hbar\omega/E_F = 0.001$. }
    \label{ground_state}
\end{figure}
We are first interested in finding the thermal equilibrium state of $S$ and $E$. We consider the case where all the impurities are prepared in the $\ket{\downarrow}_n$ internal state. In order to progress further, we will make a MF approximation to the interaction Hamiltonian 
\begin{equation}
\begin{split}
     \hat{H}_I^{\mathrm{(MF)}} =  \kappa_\downarrow\int_{-L/2}^{L/2}\!&\dd x \Big[\hat \Psi^\dagger(x)\hat{\Psi}(x)\expval{\hat n_{S}(x)}\\
     &\;+ \hat\Phi^\dagger_\downarrow(x)\hat{\Phi}_\downarrow(x)\expval{\hat n_E(x)}\Big],
\end{split}
\end{equation}
where the local densities are given by 
\begin{align}
    \expval{\hat{n}_S(x)} = \expval{\hat{\Phi}_\downarrow^\dagger(x)\hat{\Phi}_\downarrow(x)}\\
    \expval{\hat{n}_E(x)} = \expval{\hat{\Psi}^\dagger(x)\hat{\Psi}(x)}.
\end{align}
The MF approximation assumes that there are no correlations between the $S$ and $E$, which is valid for weak to moderate coupling strength \cite{Keller2022}. We can now move to a first quantization description, where the single-particle Hamiltonians acting on the particles of $S$ and $E$ are given by 
\begin{align}
    \label{h_S}
    & \hat{h}_{S} = -\frac{\hbar^2}{2M_S}\pdv[2]{x} + \kappa_\downarrow \langle\hat{n}_E(x)\rangle + V(x) \\
    \label{h_E}
    &\hat{h}_{E} = -\frac{\hbar^2}{2M_E}\pdv[2]{x} + \kappa_\downarrow \langle\hat{n}_S(x)\rangle,
\end{align}
respectively. The local average densities then take the form
\begin{align}
    \label{n_i}
    & \langle\hat{n}_S(x)\rangle = \sum_{n=1}^{N_S}|\phi_n(x)|^2\\
    \label{n_g}
    & \langle\hat{n}_E(x)\rangle = \sum_{n=1}^{N_b}f\big(\mathcal{E}_n\big
    )|\psi_n(x)|^2,
\end{align}
where $\psi_n(x)$ and $\phi_n(x)$ are the single-particle wavefunctions of $E$ and $S$. They are found by solving the single-particle Schrödinger equations 
\begin{align}
    \label{schrodinger_S}
    \hat{h}_S\phi_n(x) &= E_n\phi_n(x)\\
    \label{schrodinger_E}
    \hat{h}_E\psi_n(x) &= \mathcal{E}_n\psi_n(x)
\end{align}
self-consistently. In Eq.~(\ref{n_g}) we have assumed that the single-particle eigen-states are Fermi-Dirac distributed according to 
\begin{align}
    f(\mathcal{E}) = \frac{1}{1+e^{(\mathcal E-\mu)/k_BT}},
\end{align}
where $\mu$ is the chemical potential of the environment, found by fixing the average particle number in $E$. Furthermore, we use a basis set of size $N_b$, such that the occupation of the most energetic state $f(\mathcal{E}_{N_b})\ll 1$. We also assume that only the first $N_S$ impurity wavefunctions are occupied. This is a good approximation as long as the energy gap $E_{\mathrm{gap}} = E_{N_S+1}-E_{N_S}$ is much greater than the thermal energy $k_BT$. As we will see, this is the case when the impurities become localized and the temperature is small.

The assumption of the local density in Eq.~(\ref{n_g}) means that the state of $E$ is a thermal state $\hat{\rho}_T\propto e^{-\beta\hat{H}_E}$. Within our approximations, the full equilibrium state of $S$ and $E$ is then given by 
\begin{align}
\label{initial_state}
    \hat{\rho} = \hat{\rho}_S\otimes \hat{\rho}_T,
\end{align}
where the state of $S$ is given by 
\begin{align}
    \hat{\rho}_S = \dyad{\downarrow_1\downarrow_2\cdots\downarrow_{N_S}}\otimes \dyad{\Phi},
\end{align}
with $\ket{\Phi}$ being the $N_S$-particle state composed of the single-particle eigenstates of Eq.~(\ref{schrodinger_S}). If $N_S = 2$ the two-particle wavefunction is given by
\begin{equation}
\label{two_part_wf}
    \Phi(x_1,x_2) = \frac{1}{\sqrt{2}}\left[\phi_1(x_1)\phi_2(x_2) - \phi_1(x_2)\phi_2(x_1)\right].
\end{equation}
The state vector will then take the form 
\begin{equation}
\label{Phi_ket}
    \ket{\Phi} = \int_{-L/2}^{L/2}\dd x_1\dd x_2 \Phi(x_1,x_2)\ket{x_1,x_2}.
\end{equation}
We could instead consider the impurities to be bosons in the Tonks-Girardeau limit~\cite{tonks_complete_1936,girardeau_relationship_1960}, by adding an infinetly repulsive impurity-impurity interaction, and by imposing bosonic exchange statistics on the wavefunction. For both fermionic and Tonks-Girardeau bosons, the results presented in this manuscript are the same.

In practice Eqs.~(\ref{schrodinger_S}) and~(\ref{schrodinger_E}) are solved by starting with $\kappa_\downarrow=0$, where the problem is reduced to a simple particle in a box problem. In the case of no interactions the physical scales of the gas are fixed by the fermion number density $\Bar{n} = N_E/L$, the Fermi wavevector $k_F = \pi\Bar{n}$, the Fermi energy $E_F = \hbar^2k_F^2/2M_E$, the Fermi time $\tau_F = \hbar/E_F$, the Fermi velocity $v_F = \hbar k_F/M_E$, and the Fermi temperature $T_F = E_F/k_B$. In the rest of the paper, all quantities will be rescaled appropriately with the quantities for the case with no interaction. We then slowly increase $\kappa_\downarrow$. At each step, we use the solutions from the previous step and iterate until the solution has converged \cite{Keller2022,Keller2023,Hiyane2024}. The equilibrium densities of $E$ and $S$ are shown in Fig.~\ref{ground_state} (a-b) as the coupling strength is increased.  We can clearly see that as the interaction strength is increased, the impurities start to localize, pushing the atoms of $E$ away. Fig.~\ref{ground_state} (c) shows that as the impurities localize, a gap opens up in the spectrum. This means that our approximation of only including $N_S$ wavefunctions in the description of the impurities makes sense if the temperature is sufficiently low. Fig.~\ref{ground_state} (d) shows the single particle wavefunctions $\phi_1(x)$ and $\phi_2(x)$ at $\kappa_\downarrow k_F/E_F \approx 2.6$. Note that $\phi_1(x)$ is always symmetric and $\phi_2(x)$ anti-symmetric about the centre of the box.

Note that recent studies have shown that for impurities in a bosonic environment, the strong localization in the middle of the box is blocked due to correlations within the environment, ignored by the mean-field approximation \cite{gomezlozada2025boseferminpolaronstateemergence,breu2025}. In these studies they instead find that the impurities localize at the edge of the box potential. Due to the harmonic potential applied to the impurities, we avoid this problem.     

\section{Quench dynamics}
\label{quench_dyn}
In this section, we will investigate the quench dynamics of the densities of the impurities and the gas following a flip of the internal state of one or more of the impurities. For the sake of simplicity, we will focus on the cases $N_S = 1$ or $2$, but the generalization to more impurities is straightforward. In both cases we will consider $S$ and $E$ to have been initially prepared in the equilibrium state discussed in Sec. \ref{gs_calc} before the quench. 

\subsection{Single impurity}
\begin{figure}
    \centering
    \includegraphics[width=0.499\textwidth]{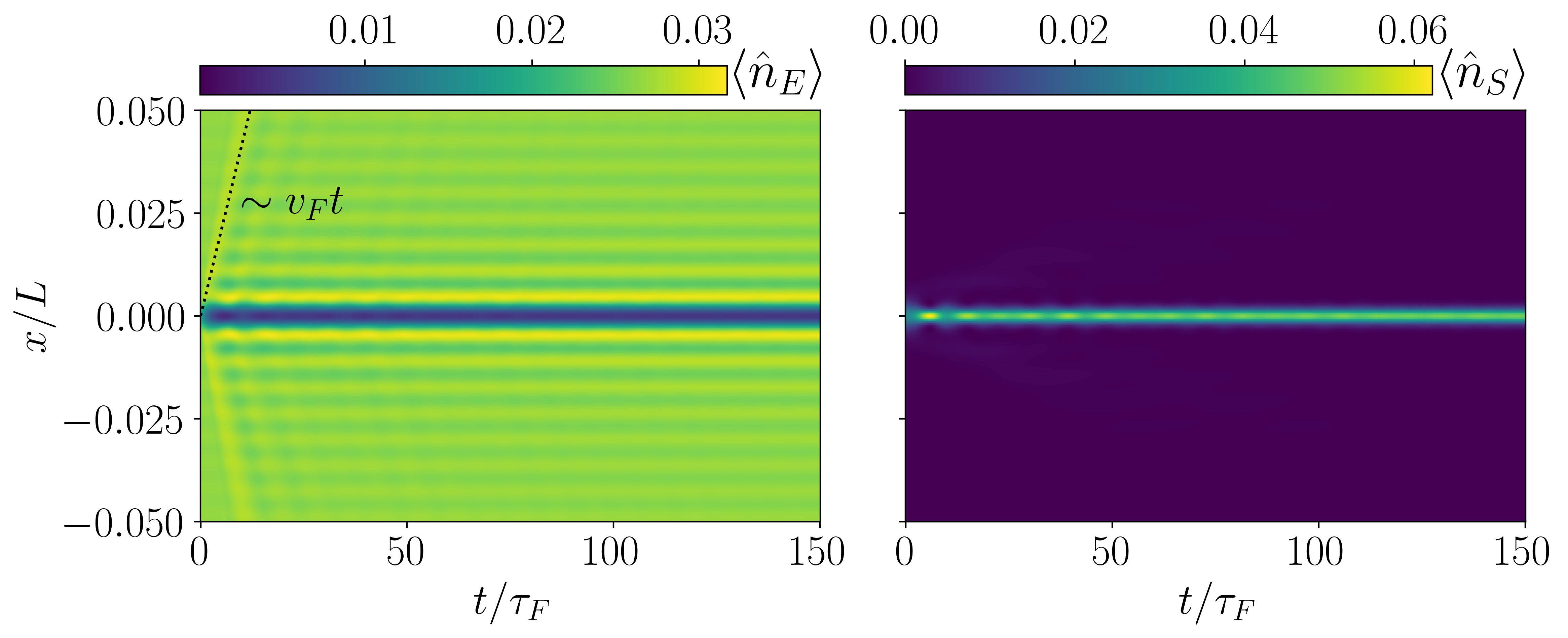}
    \caption{Quench dynamics for the density of the impurity atom (left) and the atoms of the gas (right). The parameters used are $N_E = 150$, $T = 0.005T_F$, $\kappa_\downarrow k_F/E_F \approx 1.8$, $\kappa_\uparrow = 2 \kappa_\downarrow$, $M_S/M_E=6$, and $\hbar\omega/E_F = 0.005$. }
    \label{quench_single_imp}
\end{figure}
Consider first the case of $N_S=1$. At $t=0$, the internal state is flipped $\ket{\downarrow}\rightarrow\ket{\uparrow}$. This is effectively a quench in interaction strength $\kappa_\downarrow\rightarrow\kappa_\uparrow$, such that the single-particle Hamiltonians become time-dependent
\begin{align}
    \label{h_S_t}
    & \hat{h}_{S,\uparrow}(t) = -\frac{\hbar^2}{2M_S}\pdv[2]{x} + \kappa_\uparrow \langle \hat{n}_{E,\uparrow}(x,t)\rangle + V(x) \\
    \label{h_E_t}
    &\hat{h}_{E,\uparrow}(t) = -\frac{\hbar^2}{2M_E}\pdv[2]{x} + \kappa_\uparrow\langle\hat{n}_{S,\uparrow}(x,t)\rangle,
\end{align}
where the local densities entering are now time-dependent 
\begin{align}
    \label{n_i_t}
    & \langle\hat{n}_{S,\uparrow}(x,t)\rangle = |\phi_{1,\uparrow}(x,t)|^2,\\
    & \langle\hat{n}_{E,\uparrow}(x,t)\rangle = \sum_{n=1}^{N_b}f\big(\mathcal{E}_n\big
    )|\psi_{n,\uparrow}(x,t)|^2,
\end{align}
and the single-particle wavefunctions are the solutions to the time-dependent Schrödinger equations 
\begin{align}
    & i\hbar\pdv{\phi_{n,\uparrow}(x,t)}{t} = \hat{h}_{S,\uparrow}(t)\phi_{n,\uparrow}(x,t),\\
    & i\hbar\pdv{\psi_{n,\uparrow}(x,t)}{t} = \hat{h}_{E,\uparrow}(t)\psi_{n,\uparrow}(x,t),
\end{align}
which can easily be solved numerically with high precision. 

In Fig.~\ref{quench_single_imp} we plot the density of a single impurity (left) and the non-interacting environment (right) after flipping the internal state of the impurity. For short times we see that the quench induces a breathing mode in the impurity density where the spatial distribution of the atom contracts and expands periodically. For the environment, the quench pushes the atoms close to the impurity away, creating density waves in the gas that travels with velocity $v_F$ (see black dotted line). We can also notice that density waves are excited every time the impurity distribution expands. Furthermore, the fringes observed in the gas density are Friedel oscillations~\cite{recati_casimir_2005} forming in real time with a spatial period of $\pi/2k_FL$.

\subsection{Two impurities}
\label{two_imp_quench}
\begin{figure}
    \centering
    \includegraphics[width=0.499\textwidth]{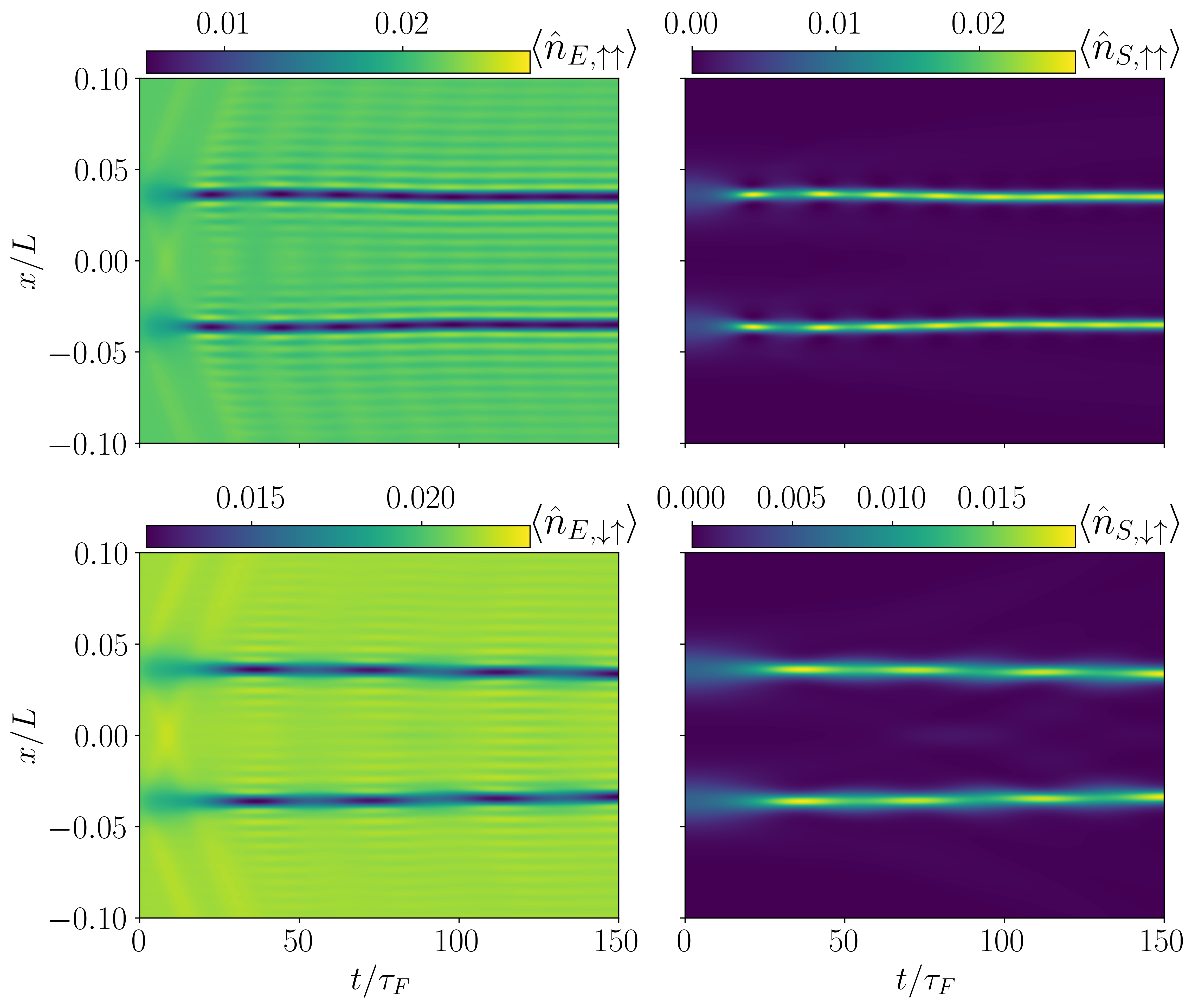}
    \caption{Quench dynamics for the density of the impurity atoms (left) and the atoms of the gas (right). The parameters used are $M_S/M_E = 2$, $N_E = 150$, $T = 0.005T_F$, $\kappa_\downarrow k_F/E_F \approx 1.8$, $\kappa_\uparrow = 2 \kappa_\downarrow$, and $\hbar\omega/E_F = 0.005$.}
    \label{quench_two_imp}
\end{figure}
We now consider the case $N_S = 2$. At time $t=0$ we flip the internal state of either one or both of the impurities. Note that flipping the internal state of only one of the impurities would be experimentally impossible since the impurities are indistinguishable. We still include the results for this case as it contributes to the decoherence dynamics studied in Sec.~\ref{decoherence_dyn}.
For $t>0$, there are four possible single-particle Hamiltonians for $E$
\begin{align}
\label{two_part_schrodinger_dd}
    \hat h_{E,\downarrow\downarrow} &= -\frac{\hbar^2}{2M_E}\pdv[2]{x} + \kappa_\downarrow(\expval{\hat{n}_{1,\downarrow}}+\expval{\hat{n}_{2,\downarrow}})\\
    \hat h_{E,\downarrow\uparrow} &= -\frac{\hbar^2}{2M_E}\pdv[2]{x} + \kappa_\downarrow\expval{\hat{n}_{1,\downarrow}}+\kappa_\uparrow\expval{\hat{n}_{2,\uparrow}}\\
    \hat h_{E,\uparrow\downarrow} &= -\frac{\hbar^2}{2M_E}\pdv[2]{x} + \kappa_\uparrow\expval{\hat{n}_{1,\uparrow}}+\kappa_\downarrow\expval{\hat{n}_{2,\downarrow}}\\
    \hat h_{E,\uparrow\uparrow}&= -\frac{\hbar^2}{2M_E}\pdv[2]{x} + \kappa_\uparrow(\expval{\hat{n}_{1,\uparrow}}+\expval{\hat{n}_{2,\uparrow}}),
    \label{two_part_schrodinger_uu}
\end{align}
We then solve the time-dependent single-particle Schrodinger equations
\begin{align}
    i\hbar\pdv{\psi_{n,\sigma}(x,t)}{t} = \hat{h}_{E,\sigma}(t)\psi_{n,\sigma}(x,t)
\end{align}
for $E$, where $\sigma$ labels the two-impurity internal state. For $S$, the situation is more complicated as the two single-particle eigenstates in general evolve with different Hamiltonians. For example, if $\sigma = \uparrow\downarrow$, the time evolution of the two single-particle wavefunctions is given by 
\begin{align}
    &i\hbar\partial_t\phi_{1,\uparrow\downarrow}(x,t) = \!\left[ -\frac{\hbar^2\partial^2_x}{2M_S} + \kappa_\uparrow\expval{\hat{n}_{E,\uparrow\downarrow}} +V(x)\right]\phi_{1,\uparrow\downarrow}(x,t)\label{pde_updown}\\
    &i\hbar\partial_t\phi_{2,\uparrow\downarrow}(x,t) = \!\left[ -\frac{\hbar^2\partial^2_x}{2M_S} + \kappa_\downarrow\expval{\hat{n}_{E,\uparrow\downarrow}}+V(x) \right]\phi_{2,\uparrow\downarrow}(x,t)\label{pde_downup},
\end{align}
and similarly for the other values of $\sigma$. This leads to the time-dependent local densities 
\begin{align}
    & \langle\hat{n}_{S,\sigma}(x,t)\rangle = \sum_{n=1}^2|\phi_{n,\sigma}(x,t)|^2\\
    & \langle\hat{n}_{E,\sigma}(x,t)\rangle = \sum_{n=1}^{N_b}f\big(\mathcal{E}_n\big
    )|\psi_{n,\sigma}(x,t)|^2,
\end{align}

Fig.~\ref{quench_two_imp} shows the density dynamics for two impurities (left) and the non-interacting gas (right) after flipping the internal state of both (top row) or one (bottom row) of the impurities. Similarly to the single impurity case, we see breathing modes in the impurity density. We can also see how the density waves induce impurity-impurity interactions mediated by the environment when the excitations travel through the gas from one impurity to the other. This induced interaction happens around a time $\tau_i = \Delta x/v_F$. Looking closely at Fig.~\ref{quench_two_imp}, we can see that around $t = \tau_i$, the density waves slightly push the impurities away from each other.

\section{Decoherence dynamics}
\label{decoherence_dyn}
\subsection{Calculating the decoherence functions}
\begin{figure*}[t]
    \centering
    \includegraphics[width=0.99\textwidth]{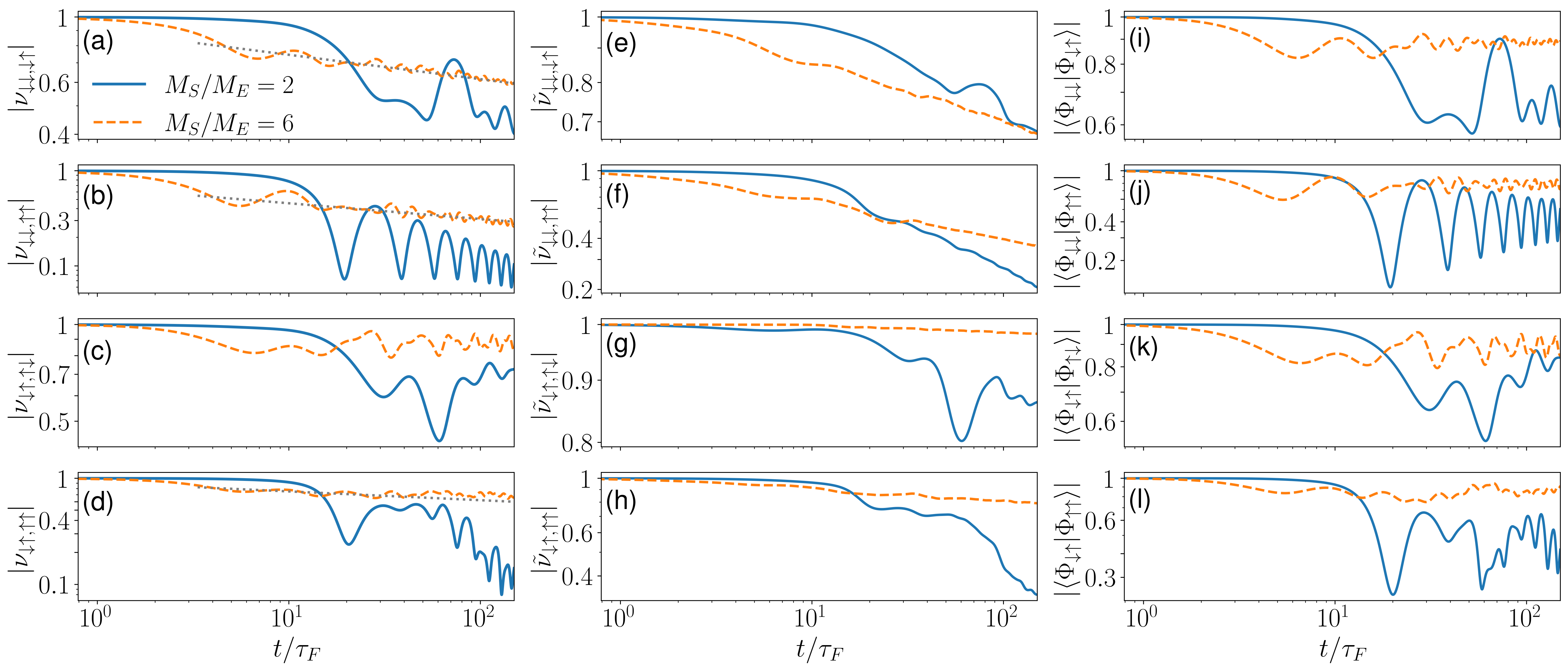}
    \caption{(a-d) Decoherence dynamics of two mobile impurities of mass $M_S = 2M_E$ (solid blue) and $M_S = 6M_E$ (dashed orange) immersed in a bath of $N_E=150$ fermions at temperature $T=0.005T_F$. The fermions in the bath are confined in a box-potential of length $k_F L = \pi N_E$, the impurity trap is set to $\hbar\omega/E_F = 0.005$ and the interaction strengths are given by $\kappa_\uparrow = 2\kappa_\downarrow$ and $\kappa_\downarrow \approx 2.1 E_F/k_F$. (e-h) The bare decoherence function (without the two-particle overlap) for the same parameters. (i-l) The two-particle wavefunction overlap of impurity wavefunctions time-evolved with different Hamiltonians.}
    \label{decoherence}
\end{figure*}
In this section, we will investigate the decoherence dynamics of two mobile impurities. We will again consider $S$ and $E$ to be prepared in the equilibrium  state discussed in Sec. \ref{gs_calc}, and at time $t=0$, the internal state is transformed into
\begin{align}
    \dyad{\downarrow_1 \downarrow_2}\longrightarrow \hat{\rho}_I(0) = \sum_{\sigma,\sigma^\prime}\rho_{\sigma,\sigma^\prime}\dyad{\sigma}{\sigma^\prime},
\end{align}
where $\sigma$ labels the different possible internal states of $S$. The total initial state of $S$ and $E$ is given by
\begin{align}
\label{rho_0}
    \hat{\rho}(0) = \sum_{\sigma,\sigma^\prime} \rho_{\sigma\sigma^\prime}\dyad{\sigma}{\sigma^\prime}\otimes\dyad{\Phi(0)}{\Phi(0)}\otimes\hat{\rho}_T
\end{align}
with $\ket{\Phi(0)}$ given by Eq.~(\ref{Phi_ket}). Due to the MF approximation, there is no scattering between internal states. Since we are starting from a statistical mixture of different internal states, each of them will thus evolve independently in time and correlations between $S$ and $E$ can build in time. Time-evolving Eq.~(\ref{rho_0}) and tracing over the environment yields the density matrix for $S$
\begin{equation}
\begin{split}
\label{rho_s_t}
    \hat\rho_S(t) &= \tr_E\hat\rho(t)\\
    &= \sum_{\sigma,\sigma^\prime}\rho_{\sigma,\sigma^\prime}\tilde\nu_{\sigma,\sigma^\prime}(t)\dyad{\sigma}{\sigma^\prime}\otimes\dyad{\Phi_\sigma(t)}{\Phi_{\sigma^\prime}(t)}.
\end{split}
\end{equation}
We refer to $\tilde\nu_{\sigma,\sigma^\prime}$ as the bare decoherence functions and they are given by 
\begin{align}
\label{bare_decoherence}
    \tilde{\nu}_{\sigma,\sigma^\prime} = \Tr \left[e^{i\hat{H}_E^{\sigma^\prime}t/\hbar}e^{-i\hat{H}_E^{\sigma}t/\hbar} \hat\rho_T\right],
\end{align}
where $\hat{H}_E^\sigma$ is the Hamiltonian for $E$ under the MF approximation, conditioned on the internal state $\sigma$ of the impurities, see App.~\ref{app_derivation_system_state} for a complete derivation.

Since Eq.~(\ref{bare_decoherence}) is a thermal expectation value of a product of exponentials of bilinear operators, we can employ the FDA~\cite{abanin_fermi-edge_2005,dambrumenil_fermi_2005,schonhammer_full_2007} to map it into a determinant in single particle space
\begin{align}
    \tilde\nu_{\sigma,\sigma^\prime}(t) = \det \left[\hat A_{\sigma,\sigma^\prime}(t)\right]
\end{align}
with matrix elements given by~(see Appendix~\ref{app:derivation} for a derivation of this)
\begin{align}
\label{fda}
    (\hat{A}_{\sigma,\sigma^\prime}(t))_{n,m} = \left[1 - f\big(\mathcal E_n\big) \right]\delta_{nm} + f\big(\mathcal E_n\big) \braket*{\psi_n^{(\sigma^\prime)}}{\psi_m^{(\sigma)}}.
\end{align}
where the single-particle overlaps of the environment wavefunctions is given by
\begin{align}
\braket*{\psi_n^{(\sigma^\prime)}}{\psi_m^{(\sigma)}} = \int_0^L\dd x \big[\psi_n^{(\sigma^\prime)}(x,t)\big]^*\psi_m^{(\sigma)}(x,t).
\end{align}
Note that $A_{\sigma,\sigma^\prime}$ in principle is an infinite dimensional operator, but since $f(\mathcal{E}_{N_b})\ll 1$, we can approximate it as a $(N_b\times N_b)$ matrix. We have checked that changing the dimension of the operator (increasing $N_b$) does not affect the results shown below.

The time-evolution of the internal state can then be found by tracing over the motional degrees of freedom of the impurities
\begin{align}
\label{rho_I}
    \hat{\rho}_I(t) = \tr_\Phi \hat\rho_S(t) = \sum_{\sigma,\sigma^\prime}\rho_{\sigma,\sigma^\prime}\nu_{\sigma,\sigma^\prime}(t)\dyad{\sigma}{\sigma^\prime},
\end{align}
where the decoherence functions are modified and are now given by
\begin{align}
    \nu_{\sigma,\sigma^\prime} = \braket{\Phi_{\sigma^\prime}(t)}{\Phi_\sigma(t)}\tilde\nu_{\sigma,\sigma^\prime}(t).
\end{align}
Thus, we can immediately see now that the decoherence dynamics of the internal state of the impurities are affected by both the motion of the impurities $\braket{\Phi_{\sigma^\prime}(t)}{\Phi_\sigma(t)}$ and the induced non-equilibrium dynamics in the environment $\tilde\nu_{\sigma,\sigma^\prime}(t)$. In this case mobile impurities can have faster decoherence rates as they deviate from their initial equilibrium positions, which can significantly affect their precision when probing the environment. 


\subsection{Results}
We will now discuss the decoherence dynamics of two impurities prepared in an equal superposition of the two internal states. Our results are shown in Fig~\ref{decoherence} for impurity mass $M_S/M_E = 2$ (solid blue lines) and $M_S/M_E = 6$ (orange dashed lines). These values are chosen as approximations of the experimentally relevant cases of Feshbach molecule impurities~\cite{regal_creation_2003,regal_lifetime_2004} and K impurities embedded in a bath of Li atoms~\cite{cetina_decoherence_2015,cetina_ultrafast_2016}, respectively. Fig.~\ref{decoherence} (a-d) shows the dynamics of representative selection of the full decoherence functions $|\nu_{\sigma,\sigma'}|$. In addition to these, the decoherence function $\nu_{\downarrow\downarrow,\uparrow\downarrow}$ is in principle different than $\nu_{\downarrow\downarrow,\downarrow\uparrow}$. However, we have checked that they behave qualitatively in the same way. Fig.~\ref{decoherence} (e-h) shows the bare decoherence functions $|\tilde\nu_{\sigma,\sigma'}|$, and (i-l) shows the contribution from the two-particle overlap $|\braket{\Phi_{\sigma^\prime}(t)}{\Phi_\sigma(t)}|$. Throughout this section, we compare our results to the case of distinguishable, infinite mass impurities studied in detail in a previous work~\cite{brattegard2024}. 

Fig.~\ref{decoherence} (a) and (b) shows the full decoherence functions $|\nu_{\downarrow\downarrow,\downarrow\uparrow}|$ and $|\nu_{\downarrow\downarrow,\uparrow\uparrow}|$ respectively, as a function of time. The first feature to note is that as the mass is increased, the decoherence functions become proportional to power laws decays
\begin{align}
    |\nu_{\downarrow\downarrow,\downarrow\uparrow}|&\propto t^{-\alpha}\\
    |\nu_{\downarrow\downarrow,\uparrow\uparrow}|&\propto t^{-2\alpha}
\end{align}
where the exponent is given by
\begin{align}
    \alpha = \left(\frac{M_\mathrm{red}(\kappa_\uparrow-\kappa_\downarrow)}{\hbar^2k_F\pi} \right)^2,
\end{align}
with $M_\mathrm{red} = \frac{M_SM_E}{M_S+M_E}$ being the reduced mass. These powerlaws are plotted as the gray dotted lines in Fig.~\ref{decoherence} (a) and (b). Remarkably, we can see the universal behavior of the Anderson orthogonality catastrophe~\cite{anderson_infrared_1967} predicted for infinite mass impurities~\cite{knap_time-dependent_2012,brattegard2024} with a relatively small mass ratio. We can see this behavior emerge from Fig.~\ref{decoherence} (e-f) and (i-j). As the mass is increased, the bare decoherence function approaches a powerlaw. Meanwhile, the two-particle overlap shows only small oscillations around a finite value for $M_S/M_E = 6$. The behavior of the overlap is more complicated for smaller mass ratios. 

Fig.~\ref{decoherence} (c) shows $|\nu_{\downarrow\uparrow,\uparrow\downarrow}|$, the corresponding bare decoherence function and two-particle overlap in panels , (g) and (k) respectively. For large impurity mass, we can see that the bare decoherence function is almost constant, and that the only source of decoherence is due to the motion of the impurities. As pointed out in~\cite{brattegard2024}, this happens because the bare decoherence function couples to density differences in the environment. Since the quench in the impurity-environment interaction strength is only able to create low energy excitations, the density difference around the two maxima in the impurity density will be minimal, and this decoherence function will only decay very slowly. For smaller impurity mass, the impurities will be less tightly localized in space, and the argument above no longer holds true. This gives rise to more complex dynamics as seen in panel (k).

Fig.~\ref{decoherence} (d) shows the dynamics of $|\nu_{\downarrow\uparrow,\uparrow\uparrow}|$, and the contribution from the bare decoherence function and two-particle overlap in panels (h) and (l) respectively. In \cite{brattegard2024} this decoherence function was predicted to have the same powerlaw evolution as $|\nu_{\downarrow\downarrow,\downarrow\uparrow}|$ for short times. For larger impurity mass, this behaviour emerges, but there seems to be some discrepency with the predicted powerlaw for long times. This might be related to the fact that second-order perturbation theory fails for this decoherence function~\cite{brattegard2024}.

One notable deviation from the prediction of~\cite{brattegard2024}, is the lack of a clear signature of the bath-induced interaction time in Fig.~\ref{decoherence} (a-d). This is because the impurities are indistinguishable, and prepared in a localized superposition of two different positions. Thus, the time it takes for the impurities to feel the presence of the other should effectively be zero. For small mass impurities there seems to be some signature of the interaction time in the bare decoherence function, e.g. in Fig.~\ref{decoherence} (e) and (g), but the interpretation of these features is less clear than that of~\cite{brattegard2024}.

\section{Thermometry with mobile impurities}
\label{thermometry}
In this section, we investigate how our proposed thermometer, consisting of one or two mobile impurities interacting with a fermionic bath can be used to estimate ultracold temperatures. We first go through a brief overview of local quantum parameter estimation theory, before explaining the various thermometry schemes we consider. Finally, we show examples of the thermometric performance of our mobile impurity thermometer. In all examples showed here we use relatively light impurities, $M_S/M_E = 2$, such that the motion of the impurities plays a significant role in the decoherence functions, as shown in Fig.~\ref{decoherence}. The initial internal state of the impurities is chosen to be $\ket{\downarrow}$ and the motional state is weakly localized inside the environment with coupling $\kappa_\downarrow \approx 1.8 E_F/k_F$. At time $t=0$ we flip the internal state(s) to an equal superposition of $\ket{\downarrow}$ and $\ket{\uparrow}$ with $\kappa_\uparrow = 2\kappa_\downarrow$ such that the quench in interaction strength has a large impact on the impurity and gas densities. 

\subsection{Local quantum thermometry}
We first review the basics of local quantum parameter estimation theory. For now we consider a general density matrix $\hat{\rho}$ that has some dependency on the temperature $T$ of the environment. By making measurements on $\hat{\rho}$ we will estimate $T$. The measurements are in general positive operator-valued measure (POVM) represented by a collection of positive operators $\hat\Pi (x)> 0$ that sum up to identity $\sum_x\hat\Pi(x) = 1$. After $M$ successive measurements, the POVM produces the outcomes $\mathbf{x} = \{x_1, x_2,\dots,x_M\}$. From the outcomes $\mathbf{x}$ we can construct a temperature estimator $\check{T}(\mathbf{x})$, e.g. using maximum-likelihood estimation. The estimator will carry uncertainty from the finite value of $M$, the number of repetitions of the experiment, as well as the random nature of quantum measurements. To quantify the precision attainable in this setup, we employ the quantum Cramer-Rao bound 
\begin{align}
    \mathbb{E}[(\check{T} - T)^2] \geq \frac{1}{M F_T}\geq \frac{1}{M\mathcal{F}_T},
\end{align}
where $F_T$ is the Fisher information for a specific measurement choice $\{\hat{\Pi}(x)\}$
\begin{align}
\label{classical_FI}
    F_T(\hat{\rho},\hat\Pi(x)) = \sum_x p(x) \left(\pdv{\ln p(x)}{x} \right)^2,
\end{align}
where $p(x) = \Tr \hat\Pi(x)\hat\rho$ is the probability of getting outcome $x$ from the measurement. The quantum Fisher information is then found by maximizing $F_T$ over all POVMs
\begin{align}
\label{qfi}
    \mathcal{F}_T = \max_{\hat\Pi(x)}F_T(\hat{\rho}_S,\hat\Pi(x)).
\end{align}
$\mathcal{F}_T$ provides the ultimate bound on the precision of the temperature estimation. The maximum of Eq. (\ref{qfi}) is reached when a projective measurement onto the eigenbasis of the symmetric-logarithmic derivative (SLD) is performed. The SLD $\hat{\Lambda}$ can be written as the solution to the operator equation   
\begin{align}
     \pdv{\hat{\rho}_S}{T} = \frac{\hat{\Lambda}\hat\rho_S + \hat{\rho}_S\hat{\Lambda}}{2}.
\end{align}
We note that the quantum Cramer-Rao bound can only be saturated in the limit of many measurements $M\gg 1$.

For a density matrix $\hat{\rho}$ that is full-rank, the QFI is given by~\cite{paris_quantum_2009}
\begin{align}
\label{qfi_full_rank}
    \mathcal{F}_T(\hat{\rho}) = 2\sum_{n,m}\frac{|\mel{r_n}{\partial_T\hat{\rho}}{r_m}|^2}{r_n+r_m},
\end{align}
where $\ket{r_n}$ and $r_n$ are the eigenvectors and eigenvalues of $\hat{\rho}$. However, if the density matrix is not full-rank we use an expression for the QFI that is defined on the support of $\hat{\rho}$ only~\cite{Liu_2014}
\begin{align}
\label{qfi_non_full_rank}
    \mathcal{F}_T(\hat{\rho}) = &\sum_{n=1}^s\left[\frac{(\partial_T r_n)}{r_n} + 4r_n\braket{\partial_T r_n}{\partial_T r_n} \right]\\ 
    -&\sum_{n,m=1}^s\frac{8 r_nr_m}{r_n+r_m}|\braket{r_n}{\partial_T r_m}|^2,
\end{align}
where $s$ is the rank of $\hat{\rho}$. Here, $r_n$ are the non-zero eigenvalues of $\hat{\rho}$ and $\ket{r_n}$ the corresponding eigenvectors. Thus, we are able to greatly simplify the calculation of the QFI if the rank of the density matrix is less than its dimension. 

We are  also interested in looking at the thermometric performance of performing sub-optimal measurements to see how the sensitivity compares to the optimal sensitivity. If we do temperature estimation by measuring the expectation value of some operator $\hat{O}$, we know that the uncertainty in the estimate is given by the error propagation formula~\cite{mehboudi_thermometry_2019}
\begin{align}
\label{error_prop_formula}
    \Delta T = \frac{\Delta \hat{O}}{\sqrt{M}\partial_T|\expval*{\hat{O}}|},
\end{align}
where $M$ is again the number of measurements we perform and $\Delta \hat{O} = \sqrt{\expval*{\hat{O^2}} - \expval*{\hat{O}}^2}$ is the operator variance of $\hat{O}$. 

As our figure of merit, we will take the sensitivity $\mathcal{S}_T(\hat{\rho},\hat{O})$. In general, we define this as 
\begin{align}
\label{sensitivity}
    \mathcal{S}_T(\hat{\rho},\hat{O}) = \frac{T}{\sqrt{M}\Delta T}.
\end{align}
The sensitivity is maximized by measurement of the SLD and can be shown to be given by $\mathcal{S}_T(\hat{\rho},\hat{\Lambda}) = T\sqrt{\mathcal{F}_T}$~\cite{mehboudi_using_2019}.

\subsubsection{Thermometry based on the full state}
\label{full_state_thermometry}
We consider first a scenario where we use the full state $\hat{\rho}_S$ of Eq.~(\ref{rho_s_t}) with the internal state prepared in an equal superposition state, e.g. all $\rho_{\sigma,\sigma^\prime} = 1/4$. We let the state evolve for some time $t$, before we make a measurement in the eigenbasis of the SLD. Because of the orthonormality of the states $\{\ket{\sigma}\otimes\ket{\Phi_\sigma}\}$, we can see that for $N_S$ impurities, the rank of the density matrix is only $s=2^{N_S}$. Since the dimension of $\hat{\rho}_S$ is in principle infinite due to the continuous wavefunction, we use Eq.~(\ref{qfi_non_full_rank}) to calculate the QFI, and the corresponding sensitivity. For the cases $N_S=1$ and $2$, we can analytically find the eigenvalues and vectors of $\hat\rho_S$ which greatly simplifies the calculation. 

To see the effect of the correlations between the impurities, we also consider the case of two independent impurities. We do this by taking the state of the two impurities to be $\hat\rho_S^{(N_S=2)} = \hat\rho_S^{(N_S=1)}\otimes\hat\rho_S^{(N_S=1)}$. Since the QFI of a product state is additive we have that $\mathcal{F}_T(\hat\rho_S^{(N_S=2)}) = 2\mathcal{F}_T(\hat\rho_S^{(N_S=1)})$.

\subsubsection{A Ramsey protocol for thermometry}
\label{internal_state_thermometry}
Next, we consider a thermometric scheme based on Ramsey interferometry, where we only make measurements on the internal state, $\hat\rho_I$ of the impurities. We prepare the same initial state as in Sec.~\ref{full_state_thermometry}, but this time we only perform the measurement on $\hat{\rho}_I(t)$ from Eq.~(\ref{rho_I}). Since this density matrix is full-rank, we use Eq.~(\ref{qfi_full_rank}) to calculate the QFI and the corresponding optimal sensitivity. 

If $N_S = 1$ the SLD will just be a linear combination of $\hat{\sigma}_x$ and $\hat\sigma_y$ which is easily measured in the lab. However, for $N_S=2$, the SLD will in general be some non-local operator which is much harder to practically implement. To this end, we are interested in sub-optimal measurements that are experimentally feasible. One such measurement is the total measurement signal from the impurities, see e.g.~\cite{adam_coherent_2022}, where after the impurities have interacted with the environment for a time $t$, a $\pi/2$ pulse is applied to the internal state with a phase $\phi$. The number of impurities left in the ground state is counted. This is equivalent to measuring the expectation value of the local operator 
\begin{align}
\label{local_obs}
    \hat{O}(\phi) = \sum_{n=1}^2 \big[\cos(\phi)\hat\sigma_n^x + \sin(\phi)\hat\sigma_n^y\big].
\end{align}
We use Eqs.~(\ref{error_prop_formula}) and~(\ref{sensitivity}) to calculate the sensitivity and numerically optimize to find the $\phi$ that yields the best sensitivity. 

\subsubsection{Thermometry based on the position distribution of impurities}
We now focus on a thermometry scheme based on the position distribution of the impurities. The first step is to trace out the internal state from $\hat{\rho}_S$, giving us the motional state
\begin{align}
\label{motional_state}
    \hat{\rho}_x = \tr_\sigma \hat{\rho}_S = \frac{1}{4}\sum_\sigma \dyad{\Phi_\sigma}.
\end{align}
Clearly, this density matrix can maximally have a rank $s=2^{N_S}$. We thus use Eq.(\ref{qfi_non_full_rank}) to calculate the QFI of $\hat\rho_x$ and the corresponding optimal sensitivity. 

However, the SLD of $\hat\rho_x$ can in general be a very complicated object that can include a joint measurement of position and momentum of the two impurities. Again, we are interested in a suboptimal measurement that can be experimentally feasible. Thus, we consider the probability distribution 
\begin{align}
\label{position_dist}
    p(x_1,x_2) = |\Phi(x_1,x_2)|^2,
\end{align}
the probability of finding one impurity at position $x_1$ and the other at $x_2$. We then use Eq.~(\ref{classical_FI}) to calculate the Fisher information of this distribution. From the Fisher information $F_T$ the sensitivity is then found as $\mathcal{S}_T = T\sqrt{F_T}$. 

\subsection{Thermometric performance}
\begin{figure}
    \centering
    \includegraphics[width=0.49\textwidth]{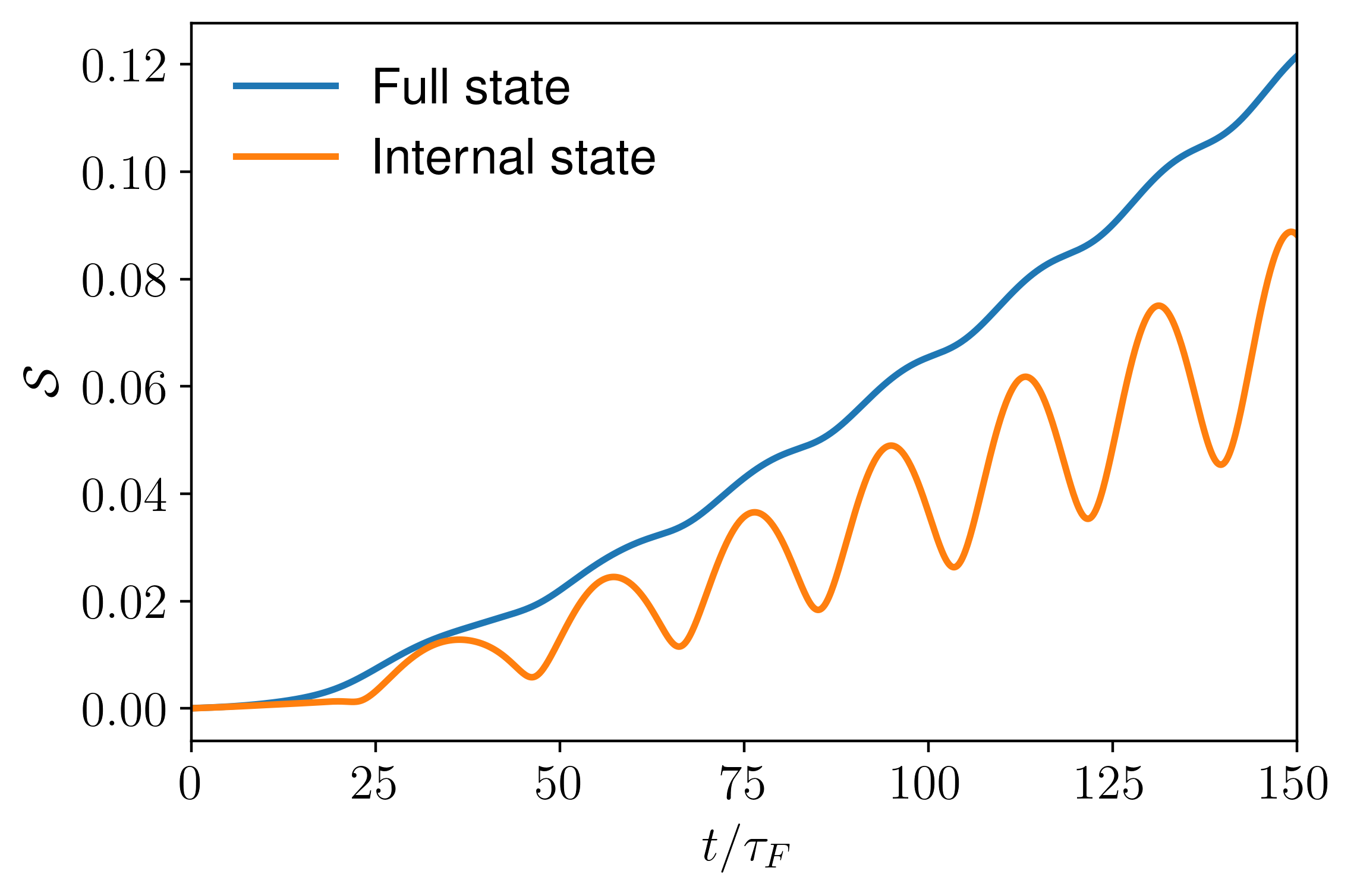}
    \caption{The optimal sensitivity as a function of time for a single impurity of mass $M_S/M_E = 2$ interacting with a bath of $N_E = 150$ fermions at temperature $T/T_F = 0.005$. The blue line is for the full state of the impurity and the orange line is for the internal state only (the spatial part has been traced out). The fermions are confined in a box-potential of length $k_F L = \pi N_E$, the impurity trap has is set by $\hbar\omega/E_F = 0.005$,  and the interaction strengths are given by $\kappa_\uparrow = 2\kappa_\downarrow$ and $\kappa_\downarrow \approx 1.8 E_F/k_F$.   }
    \label{qsnr_single_imp}
\end{figure}

We now investigate the thermometric performance of one or two mobile impurities. First we consider a single impurity interacting with a bath of cold fermions at temperature $T/T_F=0.005$. Fig.~\ref{qsnr_single_imp} shows the sensitivity of the full state (blue line) and internal state (orange line). As we can see, most of the information is captured by the internal state. However, since the impurity mass is relatively small, $M_S/M_E=2$, the breathing mode greatly impacts the coherence (see Fig.~\ref{decoherence} (a)). This is seen in the oscillations of the internal state sensitivity. Note that it seems like the sensitivity is growing linear in time. This is not true for very long times. At some point the sensitivity will reach a peak and then decrease. We are not able to simulate for these long times without running into finite size effects due to computational constraints. 
\begin{figure}
    \centering
    \includegraphics[width=0.49\textwidth]{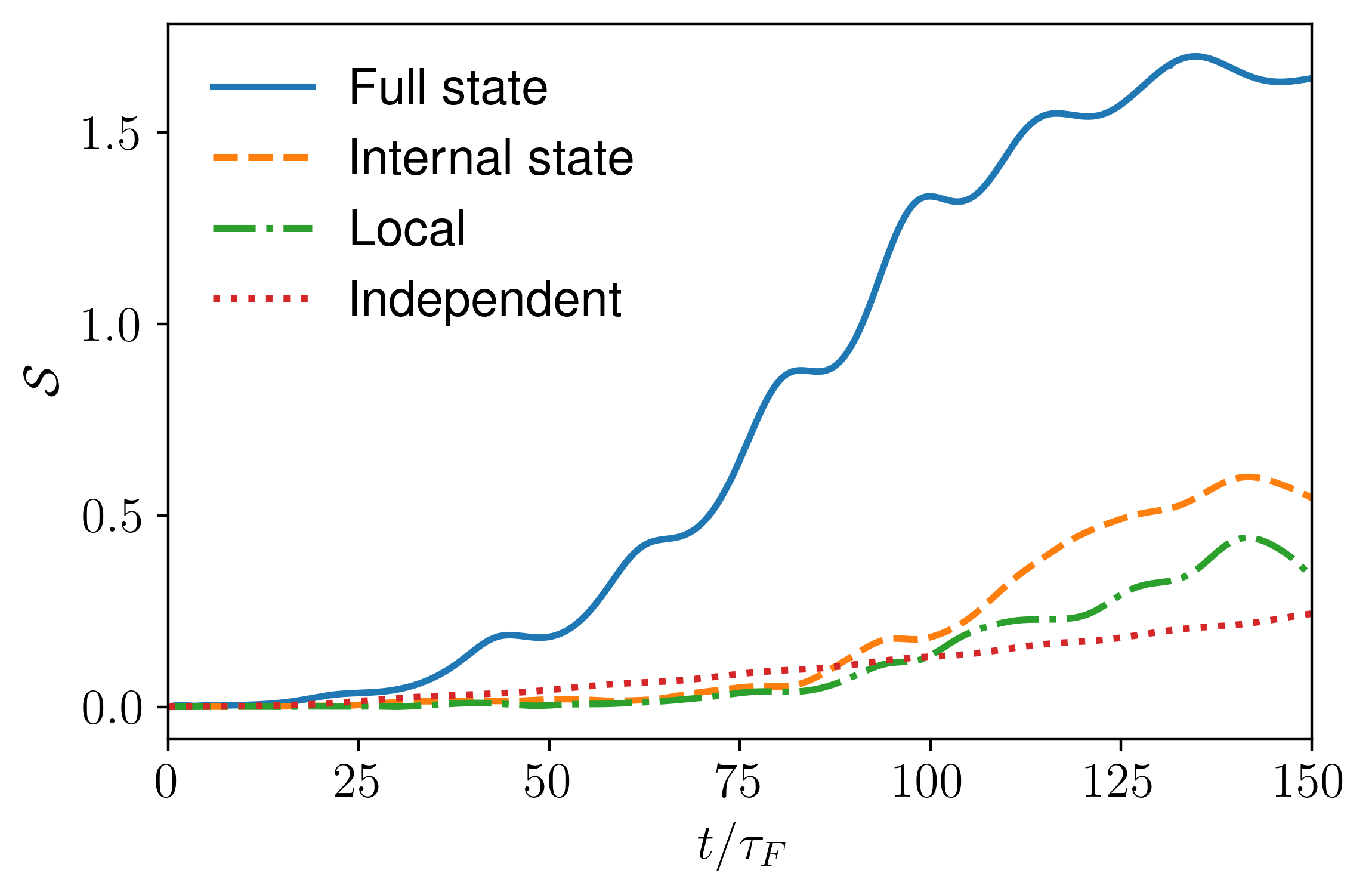}
    \caption{Sensitivity of two impurities for measuring the full state (blue line), measuring the internal state (orange dashed line), measuring the local observable of Eq.~(\ref{local_obs}) (green dashdot line), and measuring the full state of two independent impurities (red dotted line). The parameters are the same as in Fig. \ref{qsnr_single_imp}.  }
    \label{qsnr_two_imp}
\end{figure}

Next, we consider a two-impurity thermometer. We consider the sensitivity of the full state (solid blue line), the internal state (orange dashed line), and compare these to the sensitivity of the experimentally feasible measurement of the local observable of Eq.~(\ref{local_obs}) (green dashdot line). To see the effect of correlations between the impurities, we compare this with the sensitivity of two independent impurities (red dotted line). We immediately see that it is no longer true that most of the information is contained within the internal state. Instead, we see that there would be a significant thermometric advantage in accessing the full state of the impurities by measuring the SLD of the full state. We also see that there is a massive difference between the sensitivity for two correlated impurities compared to having independent impurities. It is also worth noting that measuring the local observable of Eq.~(\ref{local_obs}) can achieve a sensitivity comparable to the SLD of the internal state. Finally, we note that even though measuring the internal state is sub-optimal, it still performs better than the optimal measurement on the full state of two independent impurities.

\begin{figure}
    \centering
    \includegraphics[width=0.49\textwidth]{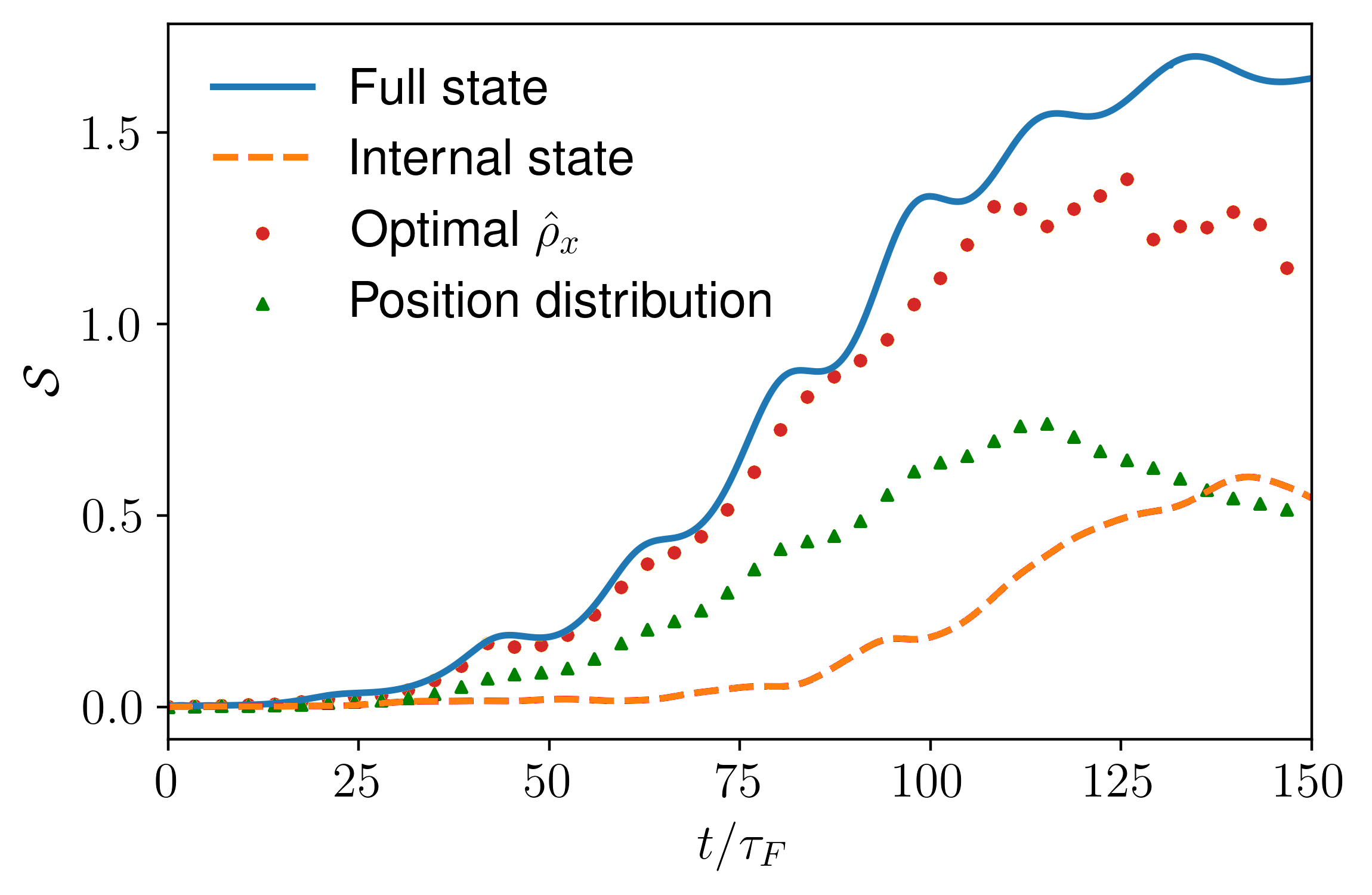}
    \caption{Sensitivity of two impurities for measuring the full state (blue line), measuring the internal state (orange dashed line), optimal measurement of the motional state of~(\ref{motional_state}) (red dots), and the position distribution of Eq. ~(\ref{position_dist}) (green triangles) The parameters are the same as in Fig. \ref{qsnr_single_imp}.  }
    \label{qsnr_two_imp_pos}
\end{figure}
In Fig.~\ref{qsnr_two_imp_pos} we compare the sensitivity of the full state (blue solid line), the internal state (orange dashed line), the motional state (red circles), and the position distribution of the impurities (green triangles). We can immediately note that the sensitivity of the motional state is almost as big as the sensitivity of the full state. However, the SLD of the motional state can in general be a complicated joint measurement of position and momentum of the impurities. We therefore look at the precision that is obtainable from the joint position distribution of the impurities. While there is a slight decrease in sensitivity as any phase information is destroyed when taking the density, we find that it still outperforms thermometry of just the internal state on short timescales. This is due to the infinite Hilbert space dimension of the continuous variable state which can potentially store more information about the thermal state of the Fermi gas than the internal state alone. 

\begin{figure}
    \centering
    \includegraphics[width=0.49\textwidth]{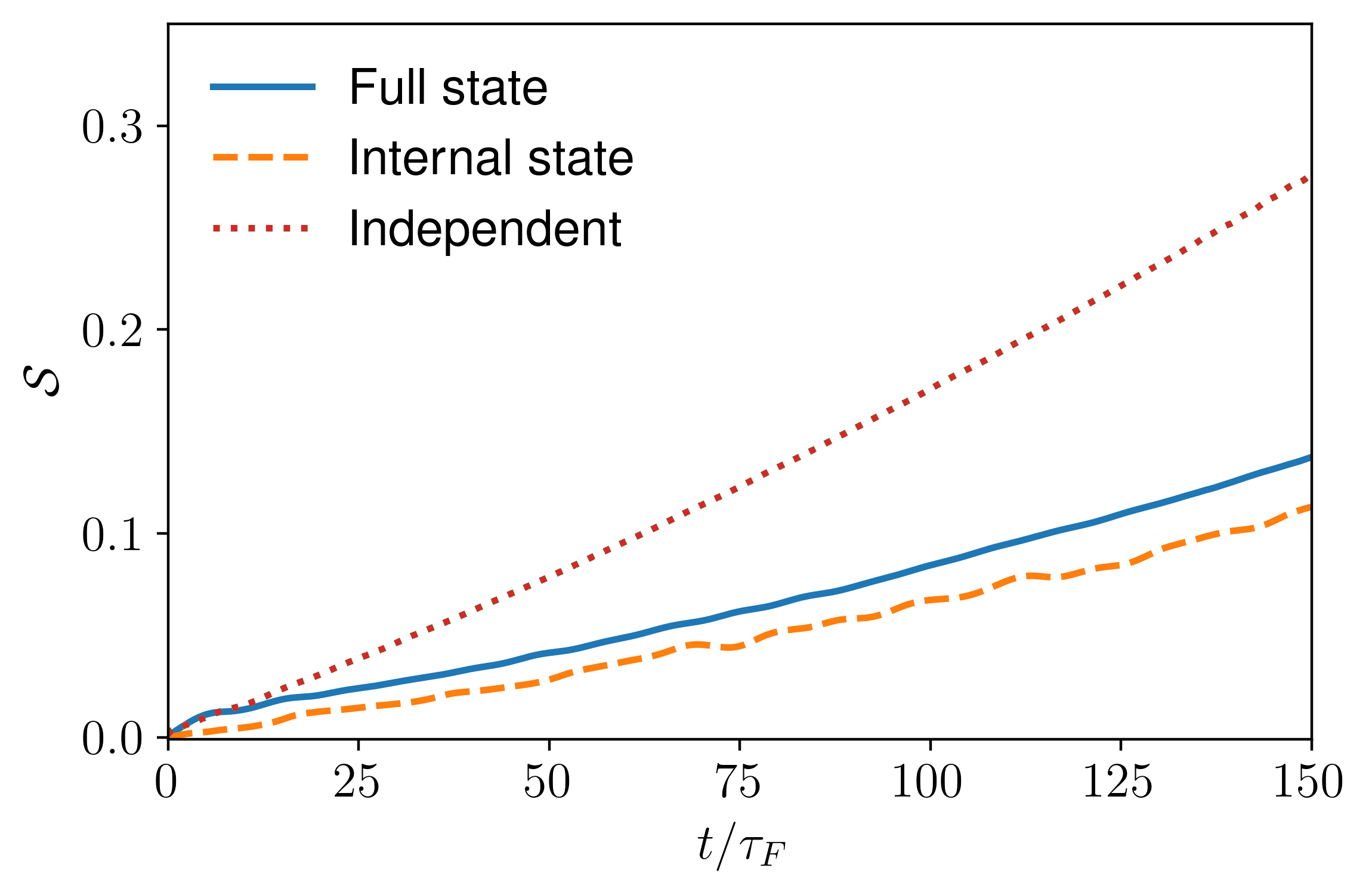}
    \caption{Sensitivity of two impurities for measuring the full state (blue line), measuring the internal state (orange dashed line), and measuring the full state of two independent impurities (red dotted line) for mass-imbalance $M_S/M_E = 6$. The other parameters are the same as in Fig. \ref{qsnr_single_imp}.  }
    \label{qsnr_two_imp_mass}
\end{figure}

Note, however, that the thermometric advantage presented above only holds for small mass imbalance. Fig.~\ref{qsnr_two_imp_mass} shows the sensitivity for two impurities, using the same parameters as Fig.~\ref{qsnr_single_imp}, except the mass imbalance is now $M_S/M_E = 6$. Clearly, the thermometric advantage vanishes in this case as the motional dynamics of the impurities are suppressed due to smaller kinetic energy, therefore, the sensitivity of the full state is mostly provided by the internal degrees of freedom. Furthermore, the collective enhancement only persists for low temperatures and when the relative coupling strength of the internal states $\kappa_\uparrow/\kappa_\downarrow$ is relatively small. For a more detailed exploration of when collective enhancement emerges, see Appendix~\ref{app_QFI}. 

\section{Discussion and conclusions}
\label{conclusion}
In this work, we have explored the equilibrium and non-equilibrium physics of mobile impurities embedded in a common bath of ultracold fermions. For a single impurity, we find that the equilibrium density becomes more localized in the center of the box as coupling strength is increased while only slightly perturbing the environment. For two impurities, the equilibrium density features two peaks, and the impurities are in a superposition of being in both locations. We then considered a quench of the interaction strength, finding that the gas is pushed away from the impurities, creating density waves travelling with the Fermi velocity, while the impurity density exhibits breathing modes. Furthermore, we studied the decoherence dynamics of two mobile impurities, and found that for sufficiently heavy impurities, we recover universal OC dynamics, as predicted for infinite mass impurities~\cite{knap_time-dependent_2012,brattegard2024}.

We also studied the effect of the motion of impurities on the thermometric performance. We find that at low temperature when the initial impurity density is weakly localized, there is significant collective enhancement in sensitivity. Remarkably, this enhancement persists even if we are restricted to experimentally relevant measurements on either the position or internal state of the impurity. This adds to the growing number of studies showing that for low temperatures there is an improvement in thermometric precision from bath-mediated correlations~\cite{planella_bath-induced_2022, brenes_multi-spin_2023,gebbia_two-qubit_2020, Xu2025, Aiache2024, brattegard2024}. We have shown that this enhancement can survive even in the presence of impurity motion. However, the role correlations play in thermometry is still not well understood. Although attempts have been made to formalize the conditions necessary for performance gain \cite{zhang2024achievingheisenbergscalinglowtemperature}, an understanding of the underlying physical mechanism for gain or loss in precision is still missing. A more complete understanding of the role of correlations for thermometry is needed and would present an interesting avenue for further research. 

This study only deals with a mean-field description of the impurity-bath coupling which limits us to only study relatively weak interactions~\cite{Keller2022}, where positional correlations can be neglected.  Our model also only holds for low temperatures such that the thermal energy is much smaller than the gap in the single particle impurity spectrum. However, we emphasise that the FDA is a non-perturbative method and the decoherence functions we compute are exact within the mean-field approximation, allowing us to capture impurity-gas correlations that are induced by the impurities' internal state. Our coupled FDA and mean-field approach could be applied to study the impact of impurity motion on other interesting problems, e.g.~decoherence in a fermionic superfluid environment~\cite{wang_exact_2022,wang_heavy_2022} or impurity-induced heating~\cite{popovic_thermodynamics_2023, Albarelli2024,Zhang2024}. To explore regions of parameter space with strong correlations beyond the mean-field approximation, other theoretical techniques such as variational methods~\cite{parish_quantum_2016, Cao2017, Liu2019, Liu2020, SuppLoc2024}, or \textit{ab initio} Hartree-Fock \cite{Theel2024} and exact diagonalization \cite{Tai2024} calculations could be employed.

\begin{acknowledgments}
We are grateful to Kesha for being very good at diagonalizing big matrices. We gratefully acknowledge the financial support of the Royal Society and Science Foundation Ireland. M.T.M. is supported by a Royal Society University Research Fellowship. This work was supported by the Okinawa Institute of Science and Technology Graduate University. This work was partially supported by Japan's Council for Science, Technology and Innovation (CSTI) under the Cross-ministerial Strategic Innovation Promotion Program (SIP) for "Promoting the application for advanced quantum technology platforms to social issues". (Funding agency: QST, Grant Number JPJ012367). T.F. acknowledges support from JSPS KAKENHI Grant No. JP23K03290. T.F. and T.B. are also supported by JST Grant No. JPMJPF2221.
\end{acknowledgments}

\appendix

\section{Derivation of Eq.~(\ref{rho_s_t})}
\label{app_derivation_system_state}
In this Appendix we will show how to arrive at the time-evolved state of $S$ and $E$. We start from the initial state
\begin{align}
    \hat{\rho}(0) = \sum_{\sigma,\sigma^\prime} \rho_{\sigma\sigma^\prime}\dyad{\sigma}{\sigma^\prime}\otimes\dyad{\Phi_\sigma(0)}{\Phi_{\sigma^\prime}(0)}\otimes\hat{\rho}_T.
\end{align}
We can write the thermal state in its eigenbasis 
\begin{align}
    \hat\rho_T = \sum_n p_n\dyad{\eta_n}.
\end{align}
Due to the MF approximation, we can write the total Hamiltonian as
\begin{align}
    \hat H = \sum_\sigma \dyad{\sigma}\otimes\left(\hat H_S^\sigma + \hat H_E^\sigma \right),
\end{align}
where $\hat{H}_S^\sigma$ and $\hat{H}_E^\sigma$ are the Hamiltonians of the system and environment under the MF approximation and conditioned on the state $\sigma$ of the internal state of the impurities. 

Consider now the unnormalized state vector
\begin{align}
    \ket{\zeta_n} = \sum_\sigma a_\sigma\ket{\sigma}\otimes\ket{\Phi_\sigma(0)}\otimes\ket{\eta_n}.
\end{align}
One can immediately notice that $\sum_n p_n\dyad{\zeta_n}{\zeta_n} = \hat{\rho}(0)$, if we define $\rho_{\sigma\sigma^\prime} = a_\sigma a_{\sigma^\prime}^*$. Now, we consider a time-evolution of this state:
\begin{equation}
\begin{split}
    e^{-i\hat H t/\hbar}\ket{\zeta_n} &= \sum_\sigma a_\sigma\ket{\sigma} e^{-i\hat H_S^\sigma t/\hbar}\ket{\Phi_\sigma(0)} e^{-i\hat H_E^\sigma t/\hbar}\ket{\eta_n}\\
    &=\sum_\sigma a_\sigma \ket{\sigma}\otimes\ket{\Phi_\sigma(t)}\otimes e^{-i\hat H_E^\sigma t/\hbar}\ket{\eta_n}.
\end{split}
\end{equation}
This means that the total time-evolved state is given by
\begin{equation}
\begin{split}
\label{app_rho_t}
    &\hat\rho(t) = e^{-i\hat H t/\hbar} \hat\rho(0)e^{i\hat H t/\hbar}\\
    &= e^{-i\hat H t/\hbar}\left[\sum_n p_n\dyad{\zeta_n}{\zeta_n}\right]e^{i\hat H t/\hbar}\\
    &=\sum_{\sigma\sigma^\prime}\rho_{\sigma\sigma^\prime}\dyad{\sigma}{\sigma^\prime}\otimes \dyad{\Phi_\sigma(t)}{\Phi_{\sigma^\prime}(t)}\otimes e^{-i\hat H_E^\sigma t/\hbar}\hat\rho_T e^{i\hat H_E^{\sigma^\prime} t/\hbar}.
\end{split}
\end{equation}
Tracing out $E$ from Eq.~(\ref{app_rho_t}) yields
\begin{align}
    \hat\rho_S(t) =\sum_{\sigma\sigma^\prime}\rho_{\sigma\sigma^\prime}\tilde{\nu}_{\sigma\sigma^\prime}\dyad{\sigma}{\sigma^\prime}\otimes \dyad{\Phi_\sigma(t)}{\Phi_{\sigma^\prime}(t)},
\end{align}
with 
\begin{align}
    \tilde{\nu}_{\sigma\sigma^\prime} = \Tr\left[e^{i\hat H_E^{\sigma^\prime} t/\hbar}e^{-i\hat H_E^\sigma t/\hbar}\hat\rho_T \right],
\end{align}
and $\ket{\Phi_\sigma(t)}$ given similarly to Eq.~(\ref{Phi_ket}), but the single-particle wavefunctions are now given by the solution to Eqs.~(\ref{pde_downup}) and~(\ref{pde_updown}), and similarly for other values of $\sigma$. We have thus arrived at Eq.~(\ref{rho_s_t}).

\section{Derivation of Eq.~(\ref{fda})}
\label{app:derivation}
We will now derive Eq.~(\ref{fda}) in the main text. According to the FDA, the expectation value of Eq.~(\ref{bare_decoherence}) can be written as a determinant in single particle space
\begin{equation}
\begin{split}
    &\Tr\left[e^{i\hat H_{\sigma^\prime }t/\hbar}e^{-i\hat H_{\sigma}t/\hbar}\hat{\rho}_T\right] = \\
    &\det\left[1-\hat n +\hat{n}e^{-i\hat h_{E,\sigma^\prime}t/\hbar}e^{i\hat h_{E,\sigma}t/\hbar}\right]
    \equiv \det \hat A_{\sigma,\sigma^\prime},
\end{split}
\end{equation}
where the occupation number operator is given by
\begin{equation}
    \hat{n} = \left[1 + e^{(\hat{h}_{E,\downarrow\downarrow} - \mu)/k_BT}\right]^{-1}.
\end{equation}
In the basis of the eigenstates $\hat{h}_{E,\downarrow\downarrow}\ket{\psi_n} = \mathcal{E}_n\ket{\psi_n}$, the matrix elements of this operator is given by
\begin{align}
    \mel{\psi_n}{\hat{A}_{\sigma,\sigma^\prime}}{\psi_m} = (1-f(\mathcal{E}_n))\delta_{nm} + f(\mathcal{E}_n)\braket*{\psi_n^{(\sigma)}}{\psi_m^{\sigma^\prime}},
\end{align}
where we have used that 
\begin{align}
    e^{i\hat h_{E,\sigma} t/\hbar}\ket{\psi_n} = \ket*{\psi_n^{(\sigma)}(t)}.
\end{align}

\section{Explorations of parameter space for thermometric performance}
\label{app_QFI}
In this Appendix, we show when a correlated impurity state fails to show collective enhancement compared to two uncorrelated impurities. As the baseline, we take the parameters used in Sec.~\ref{thermometry}, and vary one other parameter. 

\subsection{Higher temperature}
\begin{figure}[H]
    \centering
    \includegraphics[width=0.49\textwidth]{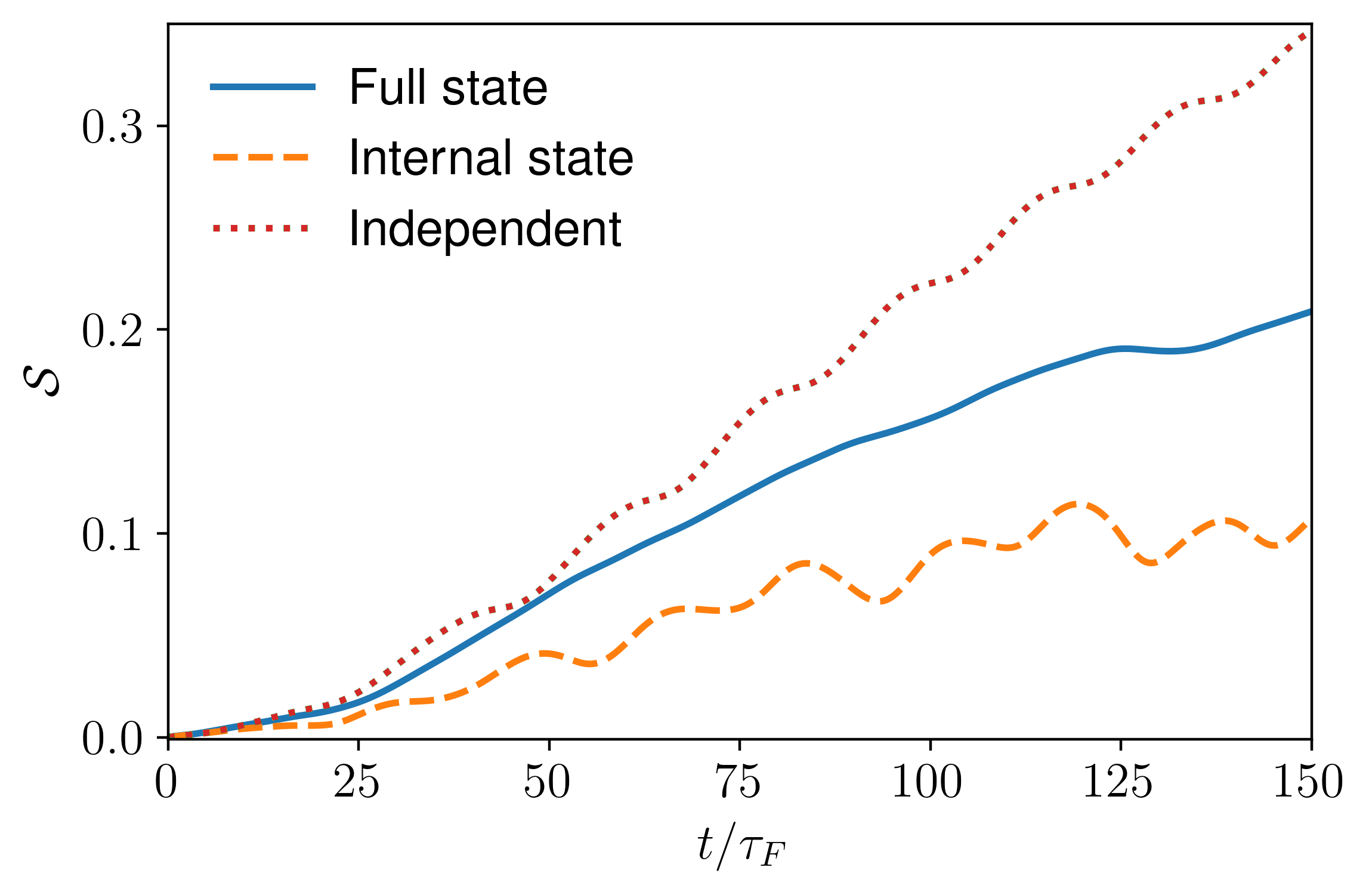}
    \caption{Sensitivity of two impurities for measuring the full state (blue line), measuring the internal state (orange dashed line),  and measuring the full state of two independent impurities (red dotted line). The temperature is $T/T_F = 0.01$ and the other parameters are the same as in Fig. \ref{qsnr_single_imp} in the main text.  }
    \label{qsnr_high_T}
\end{figure}

Figs.~\ref{qsnr_high_T} shows the sensitivity when the temperature is increased. We can clearly see that in either case the collective enhancement reported in Sec. \ref{thermometry} has vanished, and instead the independent impurities preform better as thermometers. 

\subsection{Dynamical formation of localized impurities}
\begin{figure}
    \centering
    \includegraphics[width=0.49\textwidth]{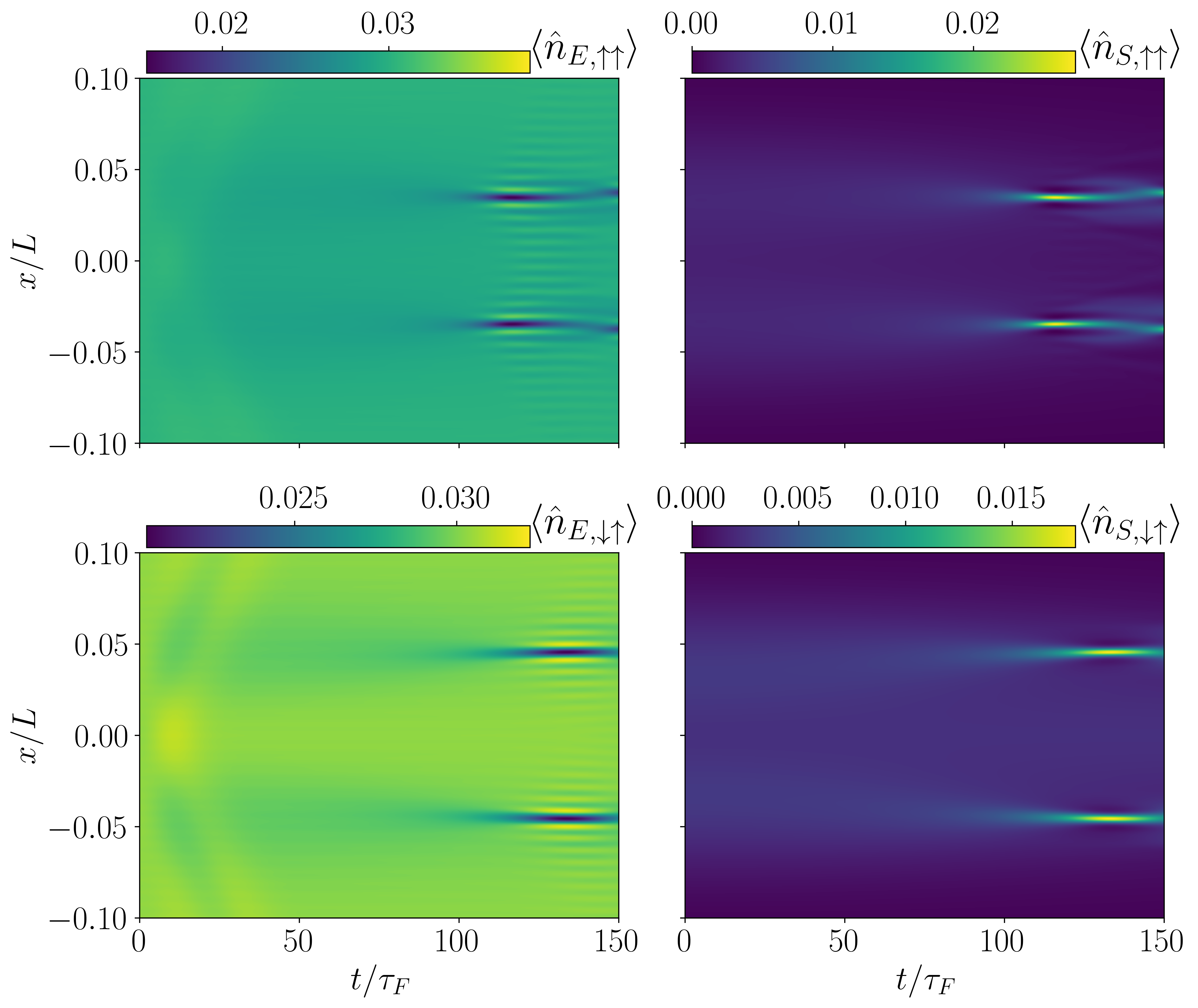}
    \caption{Same as Fig.~\ref{quench_two_imp} in the main text, but starting from no interaction, i.e. $\kappa_\downarrow=0$. The other parameters are $N_E = 150$, $T = 0.001T_F$, $\kappa_\uparrow \approx 2.7 E_F/k_F$, and $\hbar\omega/E_F = 0.005$.  }
    \label{density_no_int}
\end{figure}
\begin{figure}
    \centering
    \includegraphics[width=0.49\textwidth]{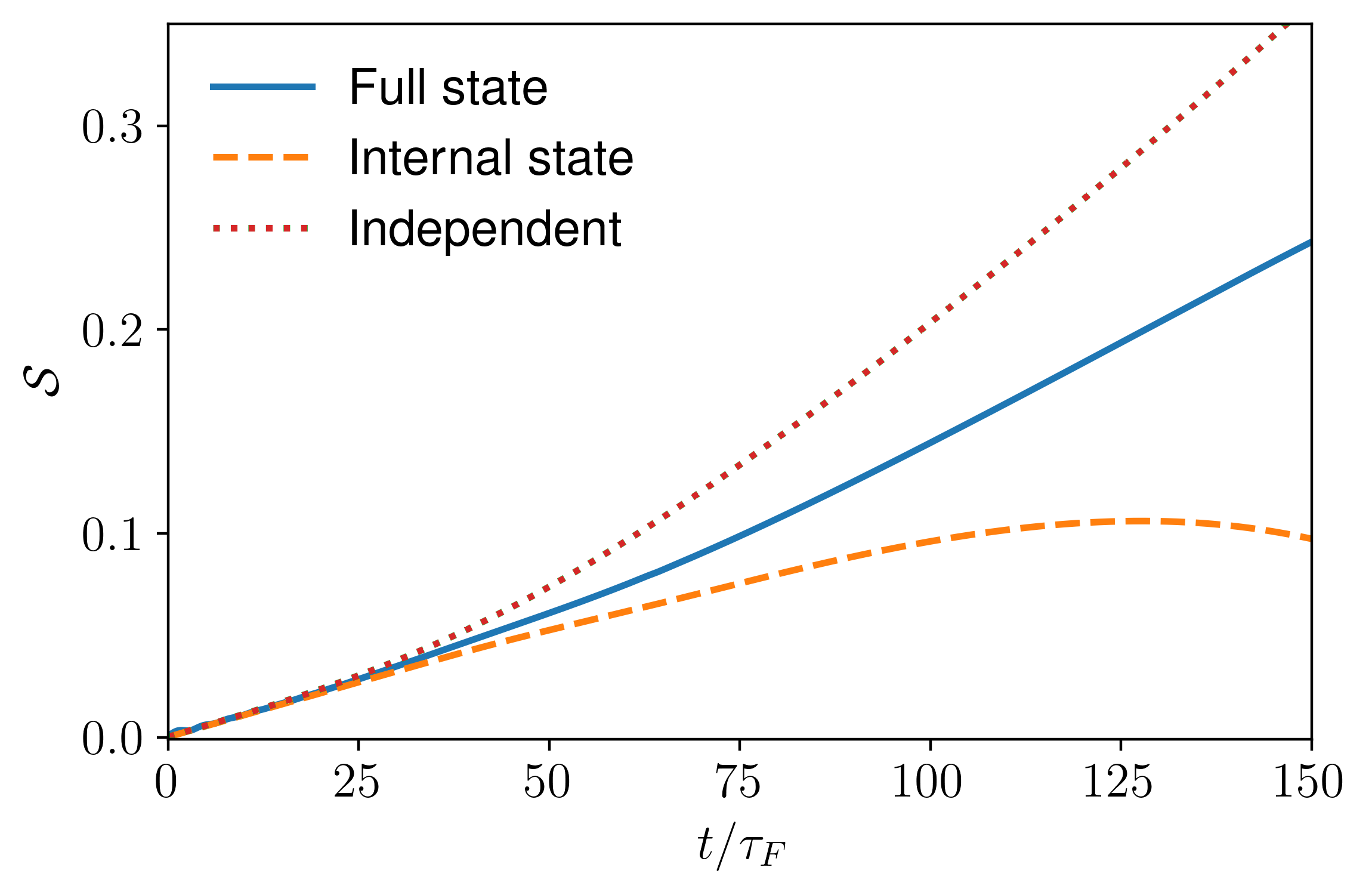}
    \caption{Same as Fig.~\ref{qsnr_high_T}, but now we start from zero interaction between impurities and gas. The mass ratio is set to $M_S/M_E = 2$, and the other parameters are the same as in Fig. \ref{density_no_int}.  }
    \label{qsnr_no_int}
\end{figure}
We now show what happens if you start from no interaction between $S$ and $E$. The initial single-particle wavefunctions of the impurity will be the eigenstates in a harmonic trap. The density evolution of the impurities if we flip the internal state of one or both of them is shown in Fig.~\ref{density_no_int}. We can clearly see that the impurities become localized in time, and exhibit breathing modes. We are interested in seeing the effect of dynamical localization on the thermometric performance. This is shown in Fig.~\ref{qsnr_no_int}. Clearly, the collective enhancement reported in the main text vanishes in this case as well.

\bibliographystyle{apsrev4-2}
\bibliography{mark_references.bib, references.bib}
\end{document}